\documentclass[lettersize,journal]{IEEEtran}

\usepackage[T1]{fontenc}      
\usepackage[utf8]{inputenc} 
\usepackage[english]{babel} 
\usepackage{mathtools}		   
\usepackage{amsfonts}			
\usepackage{amsmath}		   
\usepackage{amssymb}           
\usepackage{scalerel,stackengine} 
\usepackage{booktabs}		    
\usepackage{caption} 	           
\usepackage{multirow}		
\usepackage{tabularx}		
\usepackage{comment}		
\usepackage{url}
\usepackage{glossaries} %
\usepackage{cases}			
\usepackage{textcomp}		
\usepackage{graphicx}
\usepackage{microtype}
\usepackage[thinc]{esdiff} 	
\usepackage[normalem]{ulem} 
\usepackage{siunitx} 		
\usepackage{chemformula} 
\usepackage{diagbox} 
\usepackage{xcolor}			

\usepackage{algorithmic}
\usepackage{verbatim}
\usepackage{balance}

\usepackage[autostyle]{csquotes}			

\usepackage{keyfloat} 
\usepackage{tikz}
\usetikzlibrary{
	calc,
	positioning,
	math,
	external,
}

\usepackage{pgfplots}
\usepackage{pgfplotstable}
\usepgfplotslibrary{
}
\pgfplotsset{
	compat = newest, 
	compat = newest, 
	filter discard warning = false, 
	every axis/.append style = {
		/pgf/number format/set thousands separator = {}, 
		line width = 1.0pt, 
		line join = round, 
	},
	major tick style = { 
		thick,
		major tick length = 0.2cm,
		black,
	},
	tick align = center,
	x label style = {yshift = 0.4em}, 
	y label style = {yshift = -0.3em}, 
	legend style = {
		at = {(0.99,0.97)}, 
		font = \footnotesize, 
		fill=white!96!black,
		fill opacity = 0.5,
		text opacity = 1,
	},
	colorbar style = {
		at = {(1.03,0.5)},
		anchor = west,
		width = 8,
		tick align = inside,
	}
}

\usepackage[bookmarks=true]{hyperref} 
\RequirePackage[capitalise]{cleveref}

\newif\ifIntern 

\ifIntern
\specialcomment{intern}{\color{red}}{\color{black}}
\else
\excludecomment{intern}
\fi

\newrobustcmd{\enq}[1]{\enquote{#1}}

\newlength\figW

\NewDocumentCommand{\inputTikz}{ s t- O{1.0} m }{%
	\setlength\figW{#3\linewidth}%
	\IfBooleanF{#1}{\resizebox{#3\linewidth}{!}}
	{%
		\begingroup%
		\setlength{\fboxsep}{0pt}%
		\IfBooleanT{#2}{\fcolorbox{red}{white}}%
		{%
			\IfFileExists{#4}{%
				\tikzsetnextfilename{#4}%
				\input{#4}\unskip
			}{%
				\IfFileExists{#4.tikz}{%
					\tikzsetnextfilename{#4.tikz}%
					\input{#4.tikz}\unskip
				}{%
					\PackageError{inputTikz}{Could not find file "#4"!}{}%
				}%
			}%
		}%
		\endgroup%
	}%
}

\newcommand\diagCell[4]{%
	\multicolumn{1}{p{#2}|}{\hskip-\tabcolsep
		$\vcenter{\begin{tikzpicture}[baseline=0,anchor=south west,inner sep=#1]
				\path[use as bounding box] (0,0) rectangle (#2+2\tabcolsep,\baselineskip);
				\node[minimum width={#2+2\tabcolsep},minimum height=\baselineskip+\extrarowheight] (box) {};
				\draw (box.north west) -- (box.south east);
				\node[anchor=south west] at (box.south west) {#3};
				\node[anchor=north east] at (box.north east) {#4};
		\end{tikzpicture}}$\hskip-\tabcolsep}}
	
\newif\ifhighlight
\highlighttrue 

\newcommand{\hlbox}[1]{\ifhighlight\textcolor{black}{#1}\else#1\fi}
\stackMath
\newcommand\widecheck[1]{%
	\savestack{\tmpbox}{\stretchto{%
			\scaleto{%
				\scalerel*[\widthof{\ensuremath{#1}}]{\kern-.6pt\bigwedge\kern-.6pt}%
				{\rule[-\textheight/2]{1ex}{\textheight}}
			}{\textheight}%
		}{0.5ex}}%
	\stackon[1pt]{#1}{\scalebox{-1}{\tmpbox}}%
}

\renewcommand{\epsilon}{\varepsilon}
\renewcommand{\theta}{\vartheta}
\renewcommand{\phi}{\varphi}

\DeclarePairedDelimiter{\abs}{\lvert}{\rvert} 	
\DeclarePairedDelimiter{\norm}{\lVert}{\rVert}	

\newcommand*\dif{\mathop{}\!\mathrm{d}} 

\newcommand{\mat}[1]{\boldsymbol{#1}}	
\renewcommand{\vec}{\boldsymbol}        
\newcommand{\ver}[1]{\boldsymbol{\hat{#1}}}	

\renewcommand{\j}{\mathrm{j}}

\DeclareMathOperator{\LO}{L}  
\newcommand{\lin}[2]{\LO_{#2}{[#1]}}

\newcommand{\aeff}[2]{\mathcal{A}_{#1#2}}
\newcommand{\aeffM}[2]{\bar{\mathcal{A}}_{#1#2}}       
\newcommand{\mode}[2]{\vec{F}_{#1,#2}} 
\newcommand{\modeS}[1]{\vec{F}_{#1}} 

\newcommand{\moder}[2]{\tilde{F}_{#1,#2}} 
\newcommand{\modepol}[1]{\vec{p}_{#1}}  
\newcommand{\expp}[1]{{\operatorname{e}^{#1}}} 
\newcommand{\lag}[2]{L_{#1}^{#2}} 

\newacronym{ssmf}{SSMF}{standard single-mode fiber}
\newacronym{sdm}{SDM}{space-division multiplexing}
\newacronym{lp}{LP}{linearly polarized}
\newacronym{siso}{SISO}{single-input single-output}
\newacronym{mimo}{MIMO}{multiple-input multiple-output}
\newacronym{smf}{SMF}{single mode fiber}
\newacronym{mmf}{MMF}{multimode fiber}
\newacronym{si}{SI}{step-index}
\newacronym{gi}{GI}{graded-index}
\newacronym{gimmf}{GIMMF}{graded-index multimode fiber}
\newacronym{simmf}{SIMMF}{step-index multimode fiber}
\newacronym{mcf}{MCF}{multicore fiber}
\newacronym{gismf}{GISMF}{graded-index single mode fiber}
\newacronym{ucmcf}{UC-MCF}{uncoupled-core MCF}
\newacronym{ccmcf}{CC-MCF}{coupled-core MCF}
\newacronym{wcmcf}{WC-MCF}{weakly-coupled MCF}
\newacronym{rcmcf}{RC-MCF}{randomly-coupled MCF}
\newacronym{dsp}{DSP}{digital signal processing}
\newacronym{imgcr}{WCR}{weak coupling regime}
\newacronym{icr}{ICR}{intermediate coupling regime}
\newacronym{scr}{SCR}{strong coupling regime}
\newacronym{coi}{COI}{channel of interest}
\newacronym{smd}{SMD}{spatial mode dispersion}
\newacronym{dmgd}{DMGD}{differential mode group delay}
\newacronym{nli}{NLI}{nonlinear interference}
\newacronym{cscg}{CSCG}{circularly-symmetric complex Gaussian}
\newacronym{ase}{ASE}{amplified spontaneous emission}
\newacronym{spm}{SPM}{self-phase modulation}
\newacronym{xpm}{XPM}{cross-phase modulation}
\newacronym{dbp}{DBP}{digital back-propagation}
\newacronym{mdl}{MDL}{mode-dependent loss}
\newacronym{iir}{IIR}{intensity-impulse response}
\newacronym{gvd}{GVD}{group-velocity dispersion}
\newacronym{mbl}{MBL}{macrobend losses}
\newacronym{psd}{PSD}{power spectral density}
\newacronym{awgn}{AWGN}{additive white Gaussian noise}
\newacronym{cd}{CD}{chromatic-dispersion}
\newacronym{lma}{LMA}{large mode area fiber}
\newacronym{xt}{XT}{cross-talk}
\newacronym{air}{AIR}{achievable information rate}
\definecolor{mycolor1}{rgb}{0.12157,0.47059,0.70588}%
\definecolor{mycolor2}{rgb}{0.90588,0.16078,0.54118}%
\definecolor{mycolor3}{rgb}{0.10588,0.61961,0.46667}%
\definecolor{mycolor4}{rgb}{0.85098,0.37255,0.00784}%
\definecolor{mycolor5}{rgb}{0.46275,0.16471,0.51373}%
\definecolor{mycolor6}{rgb}{0.90196,0.67059,0.00784}%
\definecolor{mycolor7}{rgb}{0.90196,0.11765,0.07843}%
\definecolor{mycolor8}{rgb}{0.11765,1.00000,0.07843}%
\definecolor{mycolor9}{rgb}{0.11765,1.00000,1.00000}%
\definecolor{mycolor10}{rgb}{0.39216,0.73725,0.43137}%
\definecolor{mycolor11}{rgb}{0.67451,0.14118,0.12157}%
\definecolor{mycolor12}{rgb}{0.21569,0.49412,0.72157}%
\definecolor{mycolor13}{rgb}{0.30196,0.68627,0.29020}%
\definecolor{mycolor14}{rgb}{0.59608,0.30588,0.63922}%
\definecolor{mycolor15}{rgb}{1.00000,0.49804,0.00000}%
\definecolor{mycolor16}{rgb}{0.64706,0.64706,0.00000}%
\definecolor{mycolor17}{rgb}{0.00000,0.74902,0.74902}%
\definecolor{mycolor18}{rgb}{0.00000,0.00000,0.60000}%
\definecolor{mycolor19}{rgb}{0.23529,0.07059,0.29412}%
\definecolor{mycolor20}{rgb}{0.60000,0.38039,0.82745}%

\begin{document}
	\title{\hlbox{Closed-Form Expressions for Nonlinearity Coefficients in Multimode Fibers}}
	\author{Paolo Carniello, Filipe M. Ferreira,~\IEEEmembership{Senior Member,~IEEE}, Norbert Hanik~\IEEEmembership{Senior Member,~IEEE}
		\thanks{The work of Paolo Carniello was financially supported by the Federal Ministry of Education and Research of Germany in the program of ``Souver\"an. Digital. Vernetzt.''. Joint project 6G-life, project identification number: 16KISK002. The work of Filipe M. Ferreira was financially supported by a UKRI Future Leaders Fellowship [grant numbers MR/T041218/1 and MR/Y034260/1].].
		
	Paolo Carniello and Norbert Hanik are with the Insitute for Communications Engineering, Technische Universit\"at M\"unchen, Germany (email: paolo.carniello@tum.de).
	
	Filipe M. Ferreira is with the Optical Networks Group, Dept. Electronic and Electrical Eng., University College London, U.K..
}}
	
	\markboth{}%
	{}
	
	\maketitle
			
	\begin{abstract}
		\hlbox{We derive novel approximate closed-form expressions for the nonlinear coupling coefficients appearing in the Manakov equations for multimode fibers for space-division multiplexing in the two regimes of strong and weak coupling. The expressions depend only on few fiber design parameters. In particular, the Manakov coefficients are shown to be simple rational numbers which depend solely on the number of guided modes. The overall nonlinearity coefficients are found to decrease with increasing core radius and to stay nearly constant with increasing refractive index difference between core and cladding. Validation is performed through a numerical approach. The consequences of the findings onto fiber design are discussed in terms of achievable data rates. The analysis is mainly focused on the trenchless parabolic graded-index profile, but considerations on the use of realistic trenches and non-parabolic indices, and on the step-index profile are given.}
	\end{abstract}
	
	\begin{IEEEkeywords}
		Multimode fibers, space division multiplexing, Manakov equations, nonlinear coupling coefficients, optical communications.
	\end{IEEEkeywords}

\section{Introduction}\label{sec:intro}
\IEEEPARstart{I}{n} the field of \gls{sdm} the most common fiber structures are \glspl{mcf} and \glspl{mmf}. Based on the level of linear coupling among the fiber modes, two common operational regimes are distinguished: the \gls{imgcr} and the \gls{scr} \cite{sillard22}. In the \gls{imgcr}, modes belonging to the same mode group are assumed to be strongly-coupled, while modes of different groups are assumed to be uncoupled, as for \glspl{mmf} with sufficient phase-mismatch among the mode groups as well as for \glspl{wcmcf} \cite{matsui22}. In the \gls{scr}, all fiber modes are assumed to be strongly coupled, as for \glspl{rcmcf} \cite{hayashi22}, and for \glspl{gimmf} with suitable techniques to enhance coupling -- e.g., using long period fiber gratings \cite{kahn15a, arik16, liu18}. In long-haul communications it is often tried to achieve the \gls{scr} such that delay spread and \gls{mdl} accumulate at a slower pace with transmission distance (e.g., ideally with the square-root of distance), compared to the case of partial coupling \cite{ho12, antonelli15, arik15}. The level of coupling can be directly controlled in \glspl{mcf} through, e.g., the separation of the cores \cite{puttnam21, huang16}, while it is generally less straightforward to be tuned in \glspl{mmf}. Hence, \glspl{mcf} are considered to be the preferred medium for long-haul communications given their potential for smaller \gls{dsp} complexity thanks to a generally lower delay spread, when compared to \glspl{mmf} with the same number of modes \cite{rademacher22, hayashi22, puttnam21}. 

In addition, \gls{mmf}-based \gls{sdm} systems tend to have a stronger frequency-dependence in the delay spread, higher mode-averaged attenuation and \gls{mdl}, due to stronger Rayleigh scattering (because of the core's doping)  \cite{sillard22, rademacher22, ferreira24}. 

However, several \gls{mmf} system architectures have been under investigation to reduce the delay spread, from optimized fiber designs \cite{ferreira24, sillard23}, to the introduction of intentional coupling through, e.g., spinning during the manufacturing process \cite{palmieri16}, to the use of gratings as mode couplers \cite{kahn15a, arik16, liu18} or of cyclic modal permutation \cite{arik16, disciullo23}, or fiber spans with different sign of modal delay \cite{arik16, sillard22}. More importantly, \glspl{mmf} can potentially achieve a higher spatial-spectral efficiency than that of \glspl{mcf}, given their ability to support a larger number of spatial paths \cite{ferreira24, rademacher23} in the same cross-sectional area of a \gls{smf}, even above $ 1000 $, which had been speculated to be potentially needed by 2035 \cite{winzer17}. This is a strong motivation for investigating \glspl{mmf} for long distances and, thus, for bringing into play the Kerr nonlinear response of silica fibers. 

Kerr nonlinearity depends on the matrix of nonlinear coupling coefficients $ \gamma \mat{\kappa} $, which in the \gls{scr} reduces to the scalar $ \gamma \kappa $ which is sometimes assumed to scale as $ 1/M $, where $ M $ is the total number of modes \cite{antonelli16}. In such a case, under certain conditions, spectral efficiency per mode would increase with $ M $ \cite{antonelli16}. However, as we showed in \cite{carniello23}, the $ 1/M $ scaling of $ \gamma \kappa $ is valid only for a particular fiber design strategy, that is by increasing only the core radius. 

In this paper we extend the study of \cite{carniello23} by providing analytic derivations of \hlbox{approximate closed-form expressions for the nonlinearity coefficients $ \gamma $, $ \kappa $, and $ \gamma \kappa $ for the \gls{scr}, and for $ \mat{\kappa} $ and $ \gamma\mat{\kappa} $ for the \gls{imgcr}, for different \gls{gimmf} design approaches. The closed-form results clarify the relation between the nonlinearity coefficients and the fiber design parameters. In particular, we found that the elements of $ \mat{\kappa} $ are simple rational numbers which depend only on $ M $. As $ \gamma $ depends on the effective area $ \aeff{1}{1} $ of the fundamental mode, the elements of $ \gamma \mat{\kappa} $ are shown to depend only on $ \aeff{1}{1} $, once $ M $ is fixed. Maximizing $ \aeff{1}{1} $ minimizes the elements of $ \gamma \mat{\kappa} $. In particular, they decrease with increasing core radius $ R $ and stay nearly constant with increasing refractive index difference between core and cladding $ \Delta $.}

\hlbox{The derived closed-form expressions require significantly lower computational time than the current numerical approach. For example, in the case of a $ 2352 $-modes \gls{mmf}, the computation of the modal profiles with a mode solver required us $ \sim4.5 $ hours on a server equipped with two $ 16 $-cores AMD EPYC 7282 processors at $ \qty{3.2}{\giga\hertz} $ (even though no parallel workers are used in our MATLAB implementation) and $ \qty{256}{\giga\byte} $ of RAM. From the modal profiles, the computation of $ \gamma \mat{\kappa} $ from their definition required $ \sim2 $ hours on a server equipped with a $ 32 $-cores AMD EPYC 7543P processor at $ \qty{2.8}{\giga\hertz} $ (only $ 4 $ workers could be used in our MATLAB implementation not to exceed the RAM limit) and $ \qty{512}{\giga\byte} $ of RAM. On the opposite, the closed-form expressions derived here can be computed almost instantaneously.} Additionally, they free researchers who deal with, e.g., the development of models for the nonlinear \gls{sdm} channel like in \cite{serena22, lasagni23, garcia22}, from the task of choosing a specific fiber design and computing (or assuming) some values for $ \gamma \mat{\kappa} $. \hlbox{Although the analytic formulas we derive hold formally only for a trenchless \gls{gimmf} with parabolic grading, some considerations about the validity of the present results for more general fiber designs are provided in \cref{sec:comparison}, and about the step-index profile in \cref{sec:scaling}.}

The paper is organized as follows. Section \ref{sec:channelModeling} reviews the channel models for the \gls{scr} and \gls{imgcr}. Section \ref{sec:fiberDesign} reviews the basics of fiber design. Section \ref{sec:scaling} presents the main numerical and analytic results about the nonlinear coupling coefficients. In Section \ref{sec:comparison} $ \gamma \kappa $ values for optimized and manufactured fibers proposed in the literature are computed, showing agreement with our framework. Section \ref{sec:rates} \hlbox{explains the implication of this study onto fiber design through the achievable data rate in a simple \gls{sdm} scenario, highlighting that the rate does not increase in general with number of modes, but the scaling rather depends on the fiber design}. Section \ref{sec:sideRemarks} collects some side remarks. Section \ref{sec:conclusions} concludes the paper.

\section{Basics of Channel Modeling}\label{sec:channelModeling}
A common nonlinear propagation equation for the \gls{scr} is the following Manakov equation \cite{mecozzi12Scr, mumtaz13, antonelli16}
\begin{equation}\label{eq:manakovSc}
	\diffp{\vec{A}}{z} = \lin{\vec{A}}{\mathrm{scr}} - \j \gamma \kappa \norm{\vec{A}}^2 \vec{A}
\end{equation}
where $ \vec{A} = [A_1, \dots, A_M]^\mathrm{T} $ is the column vector of modal envelopes ($\mathrm{T}$ is the transpose operator), $M$ is the total number of modes (i.e., including polarizations), and $ \lin{\vec{A}}{\mathrm{scr}} $ is the linear operator accounting for all linear effects, in particular strong mode-coupling and dispersion. The last term of \cref{eq:manakovSc} is Kerr nonlinear effect, which depends on two coefficients that in the weak-guidance approximation (which is commonly assumed to hold for fibers for modern optical communications \cite{gloge71}) can be computed as \cite{mumtaz13, antonelli16}
\begin{gather}
	\gamma = \frac{\omega_0 n_2}{c \aeff{1}{1}} \label{eq:gamma} \\
	\kappa = \frac{4}{3} \frac{2N}{2N + 1} \frac{\aeff{1}{1}}{N^2} \sum_{a=1}^{N}\sum_{b=1}^{N} \frac{1}{\aeff{a}{b}} \label{eq:kSc}
\end{gather}
where $ N $ is the number of spatial modes (so that $N = M/2$), and
\begin{equation}\label{eq:aeffInter}
	\aeff{a}{b} = \frac{\int_{-\infty}^{\infty}\int_{-\infty}^{\infty} \norm{\modeS{a}}^2 \dif x \dif y \int_{-\infty}^{\infty}\int_{-\infty}^{\infty} \norm{\modeS{b}}^2 \dif x \dif y}{\int_{-\infty}^{\infty}\int_{-\infty}^{\infty} \norm{\modeS{a}}^2 \norm{\modeS{b}}^2 \dif{x}\dif y}	
\end{equation}
is the intermodal effective area between spatial modes $ \modeS{a} $ and $ \modeS{b} $ \cite[Eq. 60]{antonelli16}. The working wavelength is $ \lambda_0 = \qty{1550}{nm} $, $ \omega_0 = 2\pi c / \lambda_0 $ is the working angular frequency,  $ k_0 = \frac{2\pi}{\lambda_0} $ is the wavenumber in vacuum, the nonlinear refractive index is $ n_2 = \qty{2.6e-20}{m^{2} W^{-1}}  $, and $ c $ is the light-speed in vacuum. 

For the \gls{imgcr}, a common nonlinear propagation equation is the following Manakov equation \cite{mecozzi12Wcr}, \cite[Eq. 54]{antonelli16}
\begin{equation}\label{eq:manakovImgc}
	\diffp{\vec{A}_{a}}{z} = \lin{\vec{A}_a}{\mathrm{wcr}} - \j \gamma \sum_{b = 1}^{G} \kappa_{ab} \norm{\vec{A}_{b}}^2 \vec{A}_a
\end{equation}
where $ \vec{A}_a $ is the $ N_a $-dimensional vector obtained by stacking only the modal amplitudes of the $ a $-th mode group, $ N_x $ is the number of spatial modes of the $ x $-th mode group, $ G $ is the total number of mode groups, and this time $ \lin{\vec{A}_a}{\mathrm{wcr}} $ accounts for the linear effects of the \gls{imgcr}. 
Note that, conversely to \cref{eq:manakovSc}, \cref{eq:manakovImgc} contains multiple nonlinear coupling coefficients, which in the weak-guidance can be computed as \cite[Eq. 61 - 63]{antonelli16}
\begin{equation}\label{eq:kImgc}
	\kappa_{ab} = \frac{4}{3} \frac{2N_a}{2N_a+\delta_{ab}} \frac{\aeff{1}{1}}{\aeffM{a}{b}}
\end{equation}
where 
$ 
\delta_{ab} = 
\begin{cases}
	1 &  \text{if $ a = b $}\\
	0 &  \text{if $ a \ne b $}
\end{cases}
$ 
is the Kronecker's delta, and 
\begin{equation}\label{eq:aeffInterBar}
	\aeffM{a}{b} = \left( \frac{1}{2N_a 2N_b} \sum_{\alpha\in I_a}\sum_{\beta\in I_b} \frac{1}{\aeff{\alpha}{\beta}} \right)^{-1}
\end{equation}
with $ I_x $ being the set of indices of the modes (including polarizations) which belong to group $ x $. Note that the various $ \kappa_{ab} $ can be grouped in the matrix $ \mat{\kappa} $. In case of a single group of degenerate modes, \cref{eq:kImgc} reduces to \cref{eq:kSc}.

\hlbox{For \glspl{gimmf}, we consider only fibers supporting $ M = G (G+1) $ modes, where $G$ is the total number of mode groups, and each group consists of $ M_g = 2g $ (with $ g \in \{1,2, \dots, G\} $) modes (including polarizations), as customary in the literature \cite[Sec. 11.2.2]{agrawal21}. Then, in the \gls{imgcr} the sizes of the groups are $ M_g $. In the \gls{scr}, due to the assumption of strong linear coupling among, all modes can be thought as belonging to a single group of size $ M $.}

As a side remark, some \gls{mmf} systems might not strictly operate in either of the two considered regimes, \gls{scr} and \gls{imgcr}, but rather in an intermediate coupling regime \cite{ferreira17, ferreira19, buch19}, whose study is however outside the scope of this paper.

Throughout the paper we present and compare three methods for computing nonlinear coupling coefficients. The first method, which serves as reference approach, consists in numerically computing the modal profiles for the fiber geometry of interest with a mode solver, and then carrying out the numerical integration of \cref{eq:aeffInter} to get $\mat{\kappa}$ with \cref{eq:kImgc}. The second method consists in exploiting analytic closed-form expressions for \cref{eq:gamma} and \cref{eq:kImgc} derived later in this work. The third method consists in fitting the results obtained with the first one. In the following we will refer to the first method as the numerical method, to the second as the analytic method, and to the third as the fitted formula. 
\section{Basics of Fiber Design}\label{sec:fiberDesign}
Designing a \gls{mmf} essentially consists in selecting the refractive index profile $ n(\rho) $ of the fiber, where $ \rho $ is the radial coordinate. Common profiles, that are the ones which we consider, are the parabolic \gls{gi} and the \gls{si}, which are defined respectively as \cite[Eq. 2.78]{keiser10}
\begin{equation}\label{eq:gi}
	n_\mathrm{GI}(\rho) = 
	\begin{cases}
		n_\mathrm{core} \sqrt{1 - 2\Delta \bigl(\frac{\rho}{R}\bigr)^2}, \quad &\text{if $ \rho \le R $}\\
		n_\mathrm{core}\sqrt{1 - 2\Delta}, \quad &\text{if $ \rho > R $}
	\end{cases}
\end{equation}
and
\begin{equation}\label{key}
	n_\mathrm{SI}(\rho) = 
	\begin{cases}
		n_\mathrm{core}, \quad &\text{if $ \rho \le R $}\\
		n_\mathrm{clad}, \quad &\text{if $ \rho > R $}
	\end{cases}
\end{equation}
where $ \Delta = (n_\mathrm{core}^2 - n_\mathrm{clad}^2)/(2n_\mathrm{core}^2) $ is the refractive index difference, $ n_\mathrm{core} $ is the maximum value of $ n(\rho) $ in the core, $ n_\mathrm{clad} $ is the constant value of $ n(\rho) $ in the cladding, and $ R $ is the core radius.

In case advanced refractive index profiles (like a non-parabolic graded-index) or trenches were employed, which is not done here, few more geometrical parameters would have to be taken into account. Some considerations about the extension of the presented results to non-parabolic \gls{gi} profiles with trenches are given in \cref{sec:comparison}.

The total number of guided modes $ M $ depends on $ n(\rho) $ and on the normalized frequency \hlbox{$ V =  k_0 \, R \, \mathrm{NA} =  k_0 \, R \, n_\mathrm{core}\sqrt{2\Delta} $}, where $ \mathrm{NA} = \sqrt{n_\mathrm{core}^2 - n_\mathrm{clad}^2} = n_\mathrm{core}\sqrt{2\Delta} $ is the so-called numerical aperture \cite[Eq. 2.80]{keiser10}. For a \gls{gimmf}, an approximate relation between $ V $ and $ M $ is \cite[Eq. 2.81]{keiser10}
\begin{equation}\label{eq:nmodesgimmf}
	M \approx \frac{V^2}{4}
\end{equation}
while for a \gls{simmf} it holds \cite[Eq. 2.61]{keiser10}
\begin{equation}\label{eq:nmodessimmf}
	M \approx \frac{V^2}{2}
\end{equation}
Observe that, even though $ M $ is formally a positive integer, it will be sometimes treated here as a positive real number for simplicity of notation. 

The design of a \gls{mmf} aims at minimizing several detrimental effects like modal delays, \glspl{mdl}, and coupling losses (in particular bend losses) \cite{sillard14, ferreira14, ferreira24}, while guiding a specific number of modes. It is known that increasing $ \Delta $ tends to increase the modal delays, and the mode-averaged and relative Rayleigh scattering of the different modes, which influence the mode-averaged attenuation and the \gls{mdl} \cite{ferreira24}. 
However, a higher $ \Delta $ better confines the modal profiles, reducing the bend losses \cite{sillard17}. 
Increasing $ R $ is a straightforward approach to increase $ M $, but it is hard to support more than approximately $ 200 $ modes without increasing $ \Delta $ (beyond a conventional $\sim 0.5\%$). Hence, it is in general helpful to tune both $ R $ and $ \Delta $ when designing a fiber for a specific $ M $ \cite{sillard14, ferreira24}. As such, it is of interest to study the scaling of $ \gamma \mat{\kappa} $ in the general scenario where both $ R $ and $ \Delta $ are varied. \hlbox{Since the present work focuses on the study of the nonlinearity coefficients only, the reader is redirected to the above-referenced papers for a review and assessment of the scaling of the mentioned detrimental linear effects with fiber design parameters. The study of \gls{mmf} designs balancing linear and nonlinear effects is outside of the scope of the paper and left for future research, which can benefit from the present work. Some initial glimpses are provided in \cite{carniello24icton}.}

For the design of fibers employed later for numerical analysis, we set $ n_\mathrm{clad} \approx 1.444 $, as for a pure silica cladding at $\qty{1550}{\nm}$. 
Even though the actual fiber material is not relevant for our single-frequency study, in practice values of $ n(\rho) $ higher than $ n_\mathrm{clad} $ (e.g., for the core) can be achieved with germanium doping, while lower values (e.g., for trenches) can be obtained with fluorine doping \cite{hermann89}.

Given a fiber supporting $ M $ modes, it has always been chosen for the numerical results of this paper the highest possible $ V $ before reaching the cutoff frequency $ V_\mathrm{c} $ of the subsequent mode group, so that the guided modes exhibit the strongest confinement \cite{arik14, sillard14, ferreira24}. 

All numerical results have been obtained with the \gls{lp} mode solver implemented in MATLAB described in \cite{tasnadThesis}, based on \cite{kawano01}, and valid under the assumption of weak-guidance. \hlbox{For the \glspl{gimmf}, a square grid of $ 600\,\times\,600 $ evenly spaced points has been used, covering a square integration area with edge length of $ \qty{125}{\mu \m}$ ($ \qty{50}{\mu \m} $ for the \glspl{gimmf} with $ R = R_\mathrm{GISMF} = \qty{6.6}{\mu\meter} $). Larger grid sizes have been observed to produce negligible effects on the computation of the quantity of interest, i.e., $ \gamma \kappa_{ab} $. For the \glspl{simmf}, a square grid of at least $ 300\,\times\,300 $ evenly spaced points has been used, covering a square integration area with edge length of $ \qty{125}{\mu \m}$ ($ \qty{90}{\mu m} $ for the \glspl{gimmf} with $ R = R_\mathrm{GISMF} = \qty{4.1}{\mu\meter} $). Larger grid sizes have been observed to produce negligible effects on the computation of the quantity of interest, i.e., $ \gamma \kappa $.} The choice of modal basis is irrelevant for $ \gamma \mat{\kappa} $, as long as the modes of a mode group of a certain basis are obtained through a unitary transformation of (quasi-)degenerate vector modes \cite{schmidt17, antonelli19}. The cutoff frequencies $ V_\mathrm{c} $ for the various modes have been computed with the implicit method in \cite[Eq. 6]{lukowski77} for \glspl{gimmf}, and as zeros of Bessel functions of the first-kind for \glspl{simmf} \cite[p. 320]{snyder83}, assuming weak-guidance in both cases. 

\section{Scaling of the Nonlinear Coupling Coefficients with Index Difference and Core Radius}\label{sec:scaling}
\hlbox{This section presents the main numerical and analytical results of this work about the nonlinear coupling coefficients.} We start from the \gls{scr} in \cref{sec:scalingScr} for which the scalars $ \gamma $, $ \kappa $, and $\gamma\kappa$ are examined, followed by the \gls{imgcr} case in \cref{sec:imgcr} for which the matrices $ \mat{\kappa} $ and $ \gamma \mat{\kappa} $ have to be considered. The analysis is mainly carried out for \glspl{gimmf}; some considerations on \glspl{simmf} are provided in \cref{sec:simmf}.

\subsection{Strong Coupling Regime}\label{sec:scalingScr}
	In order to understand the scaling of $ \gamma \kappa $ in a generic scenario where both $ R $ and $ \Delta $ are varied, we firstly consider cases in which only one of the two parameters is varied, while the other is held fixed. In doing so, we also find bounds on $\gamma\kappa$, as it will be clear later. In particular, we start from a baseline \gls{gimmf} with core radius $R = R_\mathrm{GISMF} $ and refractive index $ \Delta = \Delta_\mathrm{GISMF} $. Then, a first design option is to increase only $ R $ from $ R_\mathrm{GISMF} $ to $ R_\mathrm{max} $ (fixed $\Delta)$, and then increase  $ \Delta $ from $ \Delta_\mathrm{GISMF} $ to $ \Delta_\mathrm{max} $ (fixed $R$). Starting from the same baseline  \gls{gimmf}, another design option is to increase only $ \Delta $ from $ \Delta_\mathrm{GISMF} $ to $ \Delta_\mathrm{max} $, and then increase only $R$ from $ R_\mathrm{GISMF} $ up to $ R_\mathrm{max} $. A final approach, which serves to derive a bound on $\gamma\kappa$ for \glspl{lma}, consists in starting from a baseline fiber with $ R = R_\mathrm{max} $ and $ \Delta = \Delta_\mathrm{min,GI} $, and then increasing $ \Delta $ up to $ \Delta_\mathrm{max} $.
 
    We chose $ R_\mathrm{GISMF} = \qty{6.6}{\micro \meter}$ and $ \Delta_\mathrm{GISMF} = 0.41\%$, so that $ \aeff11 \approx  \qty{86}{\micro\meter^2} $, similarly to a standard \gls{smf} \cite{corning}. We set $ \Delta_\mathrm{max} = 5\% $, $ R_\mathrm{max} = \qty{50}{\micro\meter} $, and $\Delta_\mathrm{min,GI} = 0.0072\%$. Extreme $\Delta $ and $ R $ values are considered for maximum generalization. A higher $ \Delta_\mathrm{max} $ would tend to break the weak-guidance approximation, and would not be practical since modal delays and \glspl{mdl} would be too high \cite{ferreira24}. The value of $ R_\mathrm{max} $ is bound by fixing a $ \qty{125}{\micro\meter} $ diameter cladding as for \glspl{smf} for device backward compatibility \cite{ryf20}, and mechanical reliability \cite{puttnam21, ryf20, ferreira24}. 
    
    The results of the approach in terms of $\gamma\kappa$ have been plotted in Fig.\ref{fig:giCycle}, which presents the bounds and the set of achievable values for $ \gamma\kappa $. For convenience these have been summarised in Table \ref{table:gi} for relevant fibers, with their design parameters.

    In the following section we analyse in more detail individual parts of the aforementioned approach, to derive approximate closed-form expressions for $\gamma$, $\kappa$, and $\gamma\kappa$. 

    \begin{figure}[!t]
		\centering
		\pgfplotsset{
	/pgfplots/markString/.style={
		legend image code/.code={
			\node[color=mycolor1] (n1) {\pgfuseplotmark{o}};
			\node[color=mycolor2] (n2) [right=0.1cm of n1] {\pgfuseplotmark{triangle}};
			\node[color=mycolor3] (n3) [right=0.1cm of n2] {\pgfuseplotmark{x}};
			\node[color=mycolor4] (n4) [right=0.1cm of n3] {\pgfuseplotmark{star}};
			\node[color=mycolor5] (n5) [right=0.1cm of n4] {\pgfuseplotmark{square}};
		}
	}
}
\begin{tikzpicture}
\begin{axis}[%
width=0.38\textwidth,
height=0.3\textwidth,
at={(0\textwidth,0\textwidth)},
scale only axis,
xmode=log,
xmin=1,
xmax=5e3,
xminorticks=true,
xlabel style={font=\color{white!15!black}},
xlabel={$ M $},
ymode=log,
ytick={1e-5,1e-4,1e-3},
yticklabels={$10^{-2}$, $10^{-1}$, $10^{0}$},
yminorticks=true,
ylabel style={font=\color{white!15!black}},
ylabel={$ \gamma \kappa \, (\unit{W^{-1} km^{-1}})$},
axis background/.style={fill=white},
title style={font=\bfseries},
xmajorgrids,
xminorgrids,
ymajorgrids,
yminorgrids,
legend style={legend cell align=left, align=left, draw=white!15!black},
]
\node[] at (axis cs: 2, 1.5e-3) {SMF};
\node[] at (axis cs: 2.5,	3e-5) {LMAFs};

\node[] at (axis cs: 20, 8e-4) {$ \textcolor{mycolor1}{\Delta \uparrow} $};
\node[] at (axis cs: 250, 1e-4) {$ \textcolor{mycolor2}{R \uparrow} $};
\node[] at (axis cs: 20, 1e-4) {$ \textcolor{mycolor3}{R \uparrow} $};
\node[] at (axis cs: 30, 3.5e-5) {$ \textcolor{mycolor4}{\Delta \uparrow} $};

\addlegendentry{Analytic}
\addlegendimage{dashed, color = black}

\addplot [color=mycolor1, line width=1.0pt, only marks, mark size=2.0pt, mark=o, mark options={solid, mycolor1}]
  table[row sep=crcr]{%
2	0.00109253075996758\\
6	0.00112588474072834\\
12	0.00130910497957468\\
20	0.00127974960920197\\
30	0.00134072972157825\\
};
\addlegendentry{$ R = R_\mathrm{GISMF} $}

\addplot [color=mycolor2, line width=1.0pt, only marks, mark size=2.0pt, mark=triangle, mark options={solid, mycolor2}]
  table[row sep=crcr]{%
42	0.00126239247608379\\
56	0.000949079488700412\\
72.0000000000001	0.000751635554532786\\
90.0000000000001	0.000600974781401459\\
132	0.000414141280664458\\
156	0.000353273991459229\\
182	0.000302490577743091\\
240	0.000230477013089569\\
306	0.000181387434131092\\
380	0.000146489167065511\\
462.000000000001	0.000120743428484141\\
600	9.33411926191763e-05\\
756	7.4100965395971e-05\\
992	5.65826526740731e-05\\
1190	4.72611280442224e-05\\
1560	3.60683017883019e-05\\
1892	2.97647976687396e-05\\
};
\addlegendentry{$ \Delta = \Delta_\mathrm{max} $}

\addplot [color=mycolor3, line width=1.0pt, only marks, mark size=2.0pt, mark=x, mark options={solid, mycolor3}]
  table[row sep=crcr]{%
6	0.000543086897192337\\
12	0.000292801578986757\\
20	0.000189585935085778\\
30	0.000127960816862966\\
42	9.43325128650779e-05\\
56	7.08961944724694e-05\\
72.0000000000001	5.61400363137733e-05\\
90.0000000000001	4.48892005329857e-05\\
110	3.71566717353087e-05\\
132	3.09421820693096e-05\\
156	2.6416631558884e-05\\
};
\addlegendentry{$ \Delta = \Delta_\mathrm{GISMF} $}

\addplot [color=mycolor4, line width=1.0pt, only marks, mark size=2.0pt, mark=star, mark options={solid, mycolor4}]
  table[row sep=crcr]{%
182	2.38866088080162e-05\\
210	2.37676081784769e-05\\
240	2.38876050530703e-05\\
306	2.38881182557317e-05\\
420	2.3816118978002e-05\\
552	2.38814963011336e-05\\
702	2.38353623893948e-05\\
870	2.38749327643482e-05\\
1190	2.3845956759789e-05\\
1482	2.38467632135519e-05\\
1892	2.38678932324275e-05\\
2352	2.38493265898826e-05\\
};
\addlegendentry{$ R = R_\mathrm{max} $}

\addplot [color=mycolor4, line width=1.0pt, only marks, mark size=2pt, mark=star, mark options={solid, mycolor4},forget plot]
  table[row sep=crcr]{%
2	1.91179034398599e-05\\
6	1.96963006925482e-05\\
12	2.28904558284022e-05\\
20	2.24049482976308e-05\\
30	2.35036904145707e-05\\
42	2.31503494229841e-05\\
56	2.37139488088046e-05\\
72	2.34502859510808e-05\\
90	2.38005273788947e-05\\
110	2.36140805074364e-05\\
132	2.38966750958188e-05\\
};

\addplot [color=mycolor1, dashed, line width=1.0pt]
  table[row sep=crcr]{%
2	0.000898525825212042\\
6	0.00115524748955834\\
12	0.00124411268106283\\
20	0.00128360832173149\\
30	0.00130431168175942\\
};\label{p1}

\addplot [color=mycolor2, dashed, line width=1.0pt]
table[row sep=crcr]{%
42	0.0012995155926296\\
56	0.000980336324264437\\
72	0.000765468088809218\\
90	0.000614056818495307\\
132	0.000420144138970473\\
156	0.000355918283331675\\
182	0.000305350658377448\\
240	0.000231863777938062\\
306	0.000182016841964407\\
380	0.000146664489456884\\
462	0.00012068935309519\\
600	9.29769891565273e-05\\
756	7.38166056579563e-05\\
992	5.62730820574752e-05\\
1190	4.69178593476683e-05\\
1560	3.57970342620582e-05\\
1892	2.95188433613697e-05\\
2352	2.37480537539621e-05\\
};\label{p2}

\addplot [color=mycolor3, dashed, line width=1.0pt]
table[row sep=crcr]{%
6	0.000595738217956688\\
12	0.000320782117361294\\
20	0.000198579405985563\\
30	0.000134521533086994\\
42	9.69806401324841e-05\\
56	7.31608337841547e-05\\
72	5.7125582543792e-05\\
90	4.58260167658991e-05\\
110	3.75690768080794e-05\\
132	3.1354643050352e-05\\
156	2.6561576596795e-05\\
};\label{p3}

\addplot [color=mycolor4, dashed, line width=1.0pt]
  table[row sep=crcr]{%
2	1.56559139784946e-05\\
6	2.01290322580645e-05\\
12	2.16774193548387e-05\\
20	2.23655913978495e-05\\
30	2.27263267429761e-05\\
42	2.29377344336084e-05\\
56	2.30718732314658e-05\\
72	2.31621741051701e-05\\
90	2.32258064516129e-05\\
110	2.32723045626271e-05\\
132	2.33073005093379e-05\\
2352	2.34738905720905e-05\\
};\label{p4}

\end{axis}
\end{tikzpicture}%
		\caption{Scaling of $ \gamma \kappa $ with $ M $ for different \gls{gimmf} designs. Markers indicate numerical results with the approach explained in the text. The dashed lines are the proposed theoretical trends: \hlbox{the line relative to the circles is \cref{eq:giMyFormula2} with $ R = R_\mathrm{GISMF} $; the line relative to the triangles is \cref{eq:giMyFormula} with $ \Delta = \Delta_\mathrm{max} $; the line relative to the crosses is \cref{eq:giMyFormula} with $ \Delta = \Delta_\mathrm{GISMF} $; the line relative to the stars is \cref{eq:giMyFormula2} with $ R = R_\mathrm{max} $. The symbol $R \uparrow$ indicates that only $ R $ is increased; $\Delta \uparrow$ indicates that only $ \Delta $ is increased.}\label{fig:giCycle}}
	\end{figure}

    \begin{table}[!t]
	\footnotesize
	\caption{Parameters for relevant \glspl{gimmf} in Fig.\ref{fig:giCycle}}
	\label{table:gi}
	\centering
	$ \begin{array}{c@{\hspace{1em}}c@{\hspace{1em}}c@{\hspace{1em}}c@{\hspace{0.5em}}c@{\hspace{0.5em}}c@{\hspace{1em}}}
		\toprule
		M & \Delta \, (\%) & R \, (\unit{\micro\meter}) & \aeff{1}{1} \, (\unit{\micro\meter^2}) & \kappa & \gamma \kappa \, (\unit{1/W/km}) \\
		\midrule
		2    & 0.41   & 6.6   & 86   & 8/9   & 1.1 \\
		2    & 0.0072 & 50    & 4900 & 8/9   & 0.019 \\
		42   & 5.0    & 6.8   & 22   & 0.26  & 1.3 \\
		182  & 0.43   & 50    & 574  & 0.13  & 0.024 \\
		2352 & 5.0    & 50    & 161  & 0.037 & 0.024 \\
		\bottomrule
	\end{array} $
	\vspace{-3mm}
    \end{table}
    
	\subsubsection{Scaling with the Core Radius}\label{sec:scalingR}
	To study the dependence of $ \gamma \kappa $ on $ R $, we consider the set of \glspl{gimmf} with increasing $ M $, obtained by increasing $ R $ from $ R_\mathrm{GISMF} $ to $ R_\mathrm{max} $, while keeping $ \Delta $ fixed to either $ \Delta = \Delta_\mathrm{GISMF} $ or $ \Delta = \Delta_\mathrm{max} $. We recently proposed the following approximate closed-form expressions for the fundamental mode effective area and the Manakov nonlinearity coefficient \footnote{Compared to \cite{carniello23}, a factor $ \frac{M}{M+1} $ has been prepended to the expression for $ \kappa $ to improve accuracy for small $ M $.} \cite{carniello23}
	\begin{align}
		\aeff{1}{1} &\approx \pi \frac{1}{(\mathrm{NA} k_0)^2} 4 \sqrt{M}  \label{eq:aeffApprR}\\
		\kappa &\approx \frac{M}{M+1}\frac{7}{4\sqrt{M}} \label{eq:kScPaolo}
	\end{align}
	Equation \ref{eq:aeffApprR} has been derived through the Gaussian approximation for the fundamental mode, as detailed in Appendix \ref{sec:aeff}. The comparison between the numerical result for $ \aeff{1}{1} $ and the approximate formula \cref{eq:aeffApprR} is visible in \cref{fig:giAeff} for the cases $ \Delta = \Delta_\mathrm{max} $ and $ \Delta = \Delta_\mathrm{GISMF} $.
	
	 Equation \ref{eq:kScPaolo} was obtained in \cite{carniello23} by fitting the numerical results, while in Appendix \ref{sec:k} we derive analytic results for $ \kappa $, through the infinite parabolic profile approximation. In particular, the analytic derivation yields the values of $ \kappa $ reported in \cref{tab:kSc} for any \gls{gimmf} up to $ 32 $ mode groups ($ 1056 $ modes). \hlbox{The values turn out to be rational numbers, which are independent on $ R $ or $ \Delta $ alone, but depend solely on $ M $, consistently with the numerical results.} The agreement among the analytic, the fitted and the numerical results is visible in \cref{fig:giKnl}, for the cases $ \Delta = \Delta_\mathrm{max} $ and $ \Delta = \Delta_\mathrm{GISMF} $.

	\begin{figure}[!t]
		\centering
		\begin{tikzpicture}

\begin{axis}[%
width=0.37\textwidth,
height=0.3\textwidth,
at={(0\textwidth,0\textwidth)},
scale only axis,
xmode=log,
xmin=1,
xmax=10000,
xminorticks=true,
xlabel style={font=\color{white!15!black}},
xlabel={$ M $},
ymode=log,
ytick={1e-10,1e-09},
yticklabels={$10^{2}$, $10^{3}$},
yminorticks=true,
ylabel style={font=\color{white!15!black}},
ylabel={$ \aeff{1}{1}  \, (\unit{um^2}) $},
axis background/.style={fill=white},
title style={font=\bfseries},
xmajorgrids,
xminorgrids,
ymajorgrids,
yminorgrids,
legend style={legend cell align=left, align=left, draw=white!15!black}
]

\node[] at (axis cs: 2, 1.4e-10) {SMF};
\node[] at (axis cs: 2.5,	2e-9) {LMAFs};
\node[] at (axis cs: 7, 2.5e-11) {$ \textcolor{mycolor1}{\Delta \uparrow} $};
\node[] at (axis cs: 400, 3e-11) {$ \textcolor{mycolor2}{R \uparrow} $};
\node[] at (axis cs: 40, 1.5e-10) {$ \textcolor{mycolor3}{R \uparrow} $};
\node[] at (axis cs: 6, 5e-9) {$ \textcolor{mycolor4}{\Delta \uparrow} $};
\addlegendentry{Analytic}
\addlegendimage{dashed}

\addplot [color=mycolor1, line width=1.0pt, only marks, mark size=2.0pt, mark=o, mark options={solid, mycolor1}]
  table[row sep=crcr]{%
2	8.5750194877665e-11\\
6	5.50540203059009e-11\\
12	3.6880789859877e-11\\
20	2.99689192757678e-11\\
30	2.40447698543072e-11\\
};
\addlegendentry{$ R = R_\mathrm{GISMF} $}

\addplot [color=mycolor2, line width=1.0pt, only marks, mark size=2.0pt, mark=triangle, mark options={solid, mycolor2}]
  table[row sep=crcr]{%
42	2.18261691440071e-11\\
56	2.54728183107875e-11\\
72.0000000000001	2.84604444886067e-11\\
90.0000000000001	3.20588286827082e-11\\
132	3.86437490182058e-11\\
156	4.17135102470822e-11\\
182	4.52286774838555e-11\\
240	5.18300887645041e-11\\
306	5.8431988315044e-11\\
380	6.50168846890955e-11\\
462.000000000001	7.16182758146737e-11\\
600	8.13721663714657e-11\\
756.000000000001	9.14228077206146e-11\\
992.000000000002	1.04609389838467e-10\\
1190	1.14395922762923e-10\\
1560	1.31015222811447e-10\\
1892	1.4421804141912e-10\\
};
\addlegendentry{$ \Delta = \Delta_\mathrm{max} $}

\addplot [color=mycolor3, line width=1.0pt, only marks, mark size=2.0pt, mark=x, mark options={solid, mycolor3}]
  table[row sep=crcr]{%
6	1.14198285483665e-10\\
12	1.65069577275319e-10\\
20	2.0267909160706e-10\\
30	2.52605781703133e-10\\
42	2.91962929476347e-10\\
56	3.40834506604289e-10\\
72.0000000000001	3.8086152955524e-10\\
90.0000000000001	4.29071141720692e-10\\
110	4.69761627241075e-10\\
132	5.17307715878637e-10\\
156	5.5844063024257e-10\\
};
\addlegendentry{$ \Delta = \Delta_\mathrm{GISMF} $}

\addplot [color=mycolor4, line width=1.0pt, only marks, mark size=2.0pt, mark=star, mark options={solid, mycolor4}]
  table[row sep=crcr]{%
182	5.73752964068221e-10\\
210	5.3707514509171e-10\\
240	5.00623066235106e-10\\
306	4.44028952963169e-10\\
420	3.80664439653419e-10\\
552	3.31650817831464e-10\\
702	2.9486783921459e-10\\
870	2.64670046031693e-10\\
1190	2.26745622700227e-10\\
1482	2.03279753392307e-10\\
1892	1.79851400401052e-10\\
2352	1.6148097030747e-10\\
};
\addlegendentry{$ R = R_\mathrm{max} $}

\addplot [color=mycolor4, line width=1.0pt, only marks, mark size=2pt, mark=star, mark options={solid, mycolor4},forget plot]
  table[row sep=crcr]{%
2	4.90036841406016e-09\\
6	3.15371213745574e-09\\
12	2.11177487941237e-09\\
20	1.7152913630691e-09\\
30	1.3754651348366e-09\\
42	1.18994680721375e-09\\
56	1.01927932181842e-09\\
72	9.12135995216786e-10\\
90	8.0963165822697e-10\\
110	7.39489030304059e-10\\
132	6.7161856528876e-10\\
};

\addplot [color=mycolor1, dashed, line width=1.0pt]
  table[row sep=crcr]{%
2	9.67659903930891e-11\\
30	2.49848712847801e-11\\
};\label{p5}

\addplot [color=mycolor2, dashed, line width=1.0pt]
table[row sep=crcr]{%
42	2.1391156327939e-11\\
2352	1.60076756172139e-10\\
};\label{p6}

\addplot [color=mycolor3, dashed, line width=1.0pt]
table[row sep=crcr]{%
6	1.08338218557015e-10\\
156	5.52418690490101e-10\\
};\label{p7}

\addplot [color=mycolor4, dashed, line width=1.0pt]
  table[row sep=crcr]{%
2	5.55360367269796e-09\\
2352	1.61946371806627e-10\\
};\label{p8}

\end{axis}
\end{tikzpicture}%
		\caption{Scaling of $ \aeff{1}{1} $ with $ M $ for different \gls{gimmf} designs. Markers indicate numerical results. \hlbox{The dashed lines are the proposed theoretical trends: the line relative to 
		the circles is \cref{eq:aeffApprD} with $ R = R_\mathrm{GISMF} $; the line relative to the triangles is \cref{eq:aeffApprR} with $ \Delta = \Delta_\mathrm{max} $; the line relative to crosses is \cref{eq:aeffApprR} with $ \Delta = \Delta_\mathrm{GISMF} $; the line relative to stars is \cref{eq:aeffApprD} with $ R = R_\mathrm{max} $.}\label{fig:giAeff}}
	\end{figure}
	
	\begin{figure}[!t]
		\centering
		\pgfplotsset{
	/pgfplots/markString/.style={
		legend image code/.code={
			\node[color=mycolor1] (n1) {\pgfuseplotmark{o}};
			\node[color=mycolor2] (n2) [right=0.1cm of n1] {\pgfuseplotmark{triangle}};
			\node[color=mycolor3] (n3) [right=0.1cm of n2] {\pgfuseplotmark{x}};
			\node[color=mycolor4] (n4) [right=0.1cm of n3] {\pgfuseplotmark{star}};
			\node[color=mycolor5] (n5) [right=0.1cm of n4] {\pgfuseplotmark{square}};
		}
	}
}
\begin{tikzpicture}

\begin{axis}[%
width=0.4\textwidth,
height=0.3\textwidth,
at={(0\textwidth,0\textwidth)},
scale only axis,
xmode=log,
xmin=1,
xmax=3000,
xminorticks=true,
xlabel style={font=\color{white!15!black}},
xlabel={$ M $},
ymode=log,
ymin=0.0370831726693123,
ymax=1.27279220613579,
yminorticks=true,
ylabel style={font=\color{white!15!black}},
ylabel={$ \kappa $},
axis background/.style={fill=white},
title style={font=\bfseries},
xmajorgrids,
xminorgrids,
ymajorgrids,
yminorgrids,
legend style={legend cell align=left, align=left, draw=white!15!black}
]

\addlegendimage{dashed, color = mycolor6}
\addlegendentry{Fitted}
\node[] at (axis cs: 2, 0.7) {SMF};
\addplot [color=black, only marks, mark=square, mark options={solid, black}]
table[row sep=crcr]{%
	2	0.888888888888889\\
	6	0.634920634920635\\
	12	0.478632478632479\\
	20	0.380952380952381\\
	30	0.315412186379928\\
	42	0.268733850129199\\
	56	0.233918128654971\\
	72	0.207001522070015\\
	90	0.185592185592186\\
	110	0.168168168168168\\
	132	0.153717627401838\\
	156	0.141542816702052\\
	182	0.131147540983607\\
	210	0.122169562927857\\
	240	0.114338404794836\\
	272	0.107448107448107\\
	306	0.101339124140427\\
	342	0.0958859734369938\\
	380	0.0909886264216973\\
	420	0.0865663763525996\\
	462	0.0825533957283417\\
	506	0.0788954635108481\\
	552	0.0755475185854932\\
	600	0.0724718062488445\\
	650	0.0696364567332309\\
	702	0.0670143828038565\\
	756	0.0645824159694701\\
	812	0.0623206232062321\\
	870	0.060211761704299\\
	930	0.0582408401957274\\
	992	0.0563947633434038\\
	1056	0.0546620414170083\\
};
\addlegendentry{Analytic}

\addplot [color=mycolor1, line width=1.0pt, only marks, mark size=2.0pt, mark=o, mark options={solid, mycolor1}]
  table[row sep=crcr]{%
2	0.888888465772531\\
6	0.588113913069196\\
12	0.458092488177059\\
20	0.363893727264629\\
30	0.305872437092472\\
};
\addlegendentry{$ R = R_\mathrm{GISMF}$}

\addplot [color=mycolor2, line width=1.0pt, only marks, mark size=2.0pt, mark=triangle, mark options={solid, mycolor2}]
  table[row sep=crcr]{%
42	0.261426973870097\\
56	0.229381330451871\\
72	0.202967954829495\\
90	0.182802605178994\\
132	0.15184701423026\\
156	0.139819223179324\\
182	0.129808825988995\\
240	0.11334126394512\\
306	0.100562565408173\\
380	0.0903670588635396\\
462	0.0820475933224465\\
600	0.0720655500469792\\
756	0.0642771930108595\\
992	0.0561606925564147\\
1190	0.0512971351198706\\
1560	0.0448359046009863\\
1892	0.0407287432824959\\
};
\addlegendentry{$ \Delta = \Delta_\mathrm{max} $}

\addplot [color=mycolor3, line width=1.0pt, only marks, mark size=2.0pt, mark=x, mark options={solid, mycolor3}]
  table[row sep=crcr]{%
6	0.588447050578288\\
12	0.458584039318499\\
20	0.364580592121543\\
30	0.30668940449596\\
42	0.26131696041728\\
56	0.229268802592107\\
72	0.202870209776362\\
90	0.182746748492321\\
110	0.165612390433492\\
132	0.151872231871126\\
156	0.139969344711235\\
};
\addlegendentry{$ \Delta = \Delta_\mathrm{GISMF} $}

\addplot [color=mycolor4, line width=1.0pt, only marks, mark size=2.0pt, mark=star, mark options={solid, mycolor4}]
  table[row sep=crcr]{%
182	0.130034299112644\\
210	0.121115301760827\\
240	0.113465007905707\\
306	0.100640251040715\\
420	0.0860184834632391\\
552	0.0751486336317507\\
702	0.0666849221351469\\
870	0.0599550034029277\\
1190	0.0513017459577276\\
1482	0.0459940916650891\\
1892	0.040729248043328\\
2352	0.0365406234532744\\
};
\addlegendentry{$ R = R_\mathrm{max} $}

\addplot [color=mycolor4, line width=1.0pt, only marks, mark size=2pt, mark=star, mark options={solid, mycolor4},forget plot]
  table[row sep=crcr]{%
2	0.888888888776693\\
6	0.589366161468341\\
12	0.458649096522242\\
20	0.364636659105082\\
30	0.30673555960086\\
42	0.261374719890156\\
56	0.2293377638432\\
72	0.20294867446258\\
90	0.18283214052745\\
110	0.165684261892178\\
132	0.15227852214645\\
};

\addplot [color=mycolor6, dashed, line width=1.0pt, forget plot]
  table[row sep=crcr]{%
2 0.824958\\
3 0.757772\\
4 0.700000\\
5 0.652186\\
6 0.612372\\
7 0.578758\\
8 0.549972\\
9 0.525000\\
10 0.503090\\
11 0.483674\\
12 0.466321\\
13 0.450694\\
14 0.436527\\
15 0.423608\\
16 0.411765\\
17 0.400857\\
18 0.390770\\
19 0.381404\\
20 0.372678\\
21 0.364523\\
22 0.356879\\
23 0.349696\\
24 0.342929\\
25 0.336538\\
26 0.330492\\
27 0.324760\\
28 0.319315\\
29 0.314135\\
30 0.309198\\
31 0.304487\\
32 0.299985\\
33 0.295676\\
34 0.291548\\
35 0.287587\\
36 0.283784\\
37 0.280127\\
38 0.276608\\
39 0.273219\\
40 0.269951\\
41 0.266797\\
42 0.263751\\
43 0.260807\\
44 0.257960\\
45 0.255203\\
46 0.252534\\
47 0.249946\\
48 0.247436\\
49 0.245000\\
50 0.242635\\
51 0.240337\\
52 0.238102\\
53 0.235929\\
54 0.233815\\
55 0.231756\\
56 0.229751\\
57 0.227797\\
58 0.225892\\
59 0.224033\\
60 0.222220\\
61 0.220451\\
62 0.218722\\
63 0.217034\\
64 0.215385\\
65 0.213772\\
66 0.212195\\
67 0.210652\\
68 0.209143\\
69 0.207666\\
70 0.206219\\
71 0.204802\\
72 0.203414\\
73 0.202054\\
74 0.200721\\
75 0.199414\\
76 0.198132\\
77 0.196874\\
78 0.195640\\
79 0.194429\\
80 0.193240\\
81 0.192073\\
82 0.190927\\
83 0.189801\\
84 0.188694\\
85 0.187607\\
86 0.186538\\
87 0.185488\\
88 0.184455\\
89 0.183439\\
90 0.182439\\
91 0.181456\\
92 0.180488\\
93 0.179536\\
94 0.178599\\
95 0.177676\\
96 0.176767\\
97 0.175872\\
98 0.174991\\
99 0.174123\\
100 0.173267\\
101 0.172424\\
122 0.157149\\
152 0.141016\\
182 0.129010\\
212 0.119626\\
242 0.112031\\
272 0.105721\\
302 0.100369\\
332 0.095755\\
362 0.091725\\
392 0.088163\\
422 0.084987\\
452 0.082131\\
482 0.079545\\
512 0.077189\\
542 0.075031\\
572 0.073043\\
602 0.071206\\
632 0.069501\\
662 0.067913\\
692 0.066429\\
722 0.065038\\
752 0.063731\\
782 0.062500\\
812 0.061337\\
842 0.060237\\
872 0.059195\\
902 0.058204\\
932 0.057262\\
962 0.056364\\
992 0.055507\\
1022 0.054687\\
1052 0.053904\\
1082 0.053152\\
1112 0.052432\\
1142 0.051740\\
1172 0.051074\\
1202 0.050434\\
1232 0.049817\\
1262 0.049223\\
1292 0.048649\\
1322 0.048094\\
1352 0.047559\\
1382 0.047040\\
1412 0.046539\\
1442 0.046053\\
1472 0.045582\\
1502 0.045125\\
1532 0.044681\\
1562 0.044251\\
1592 0.043832\\
1622 0.043426\\
1652 0.043030\\
1682 0.042645\\
1712 0.042270\\
1742 0.041905\\
1772 0.041549\\
1802 0.041202\\
1832 0.040864\\
1862 0.040534\\
1892 0.040211\\
1922 0.039897\\
1952 0.039589\\
1982 0.039289\\
2012 0.038995\\
2042 0.038708\\
2072 0.038427\\
2102 0.038152\\
2132 0.037883\\
2162 0.037619\\
2192 0.037361\\
2222 0.037108\\
2252 0.036860\\
2282 0.036618\\
2312 0.036379\\
2342 0.036146\\
};
\end{axis}
\end{tikzpicture}%
		\caption{Scaling of $ \kappa $ with $ M $ for different \gls{gimmf} designs. Black squares indicate analytic results (from \cref{tab:kSc}). The other markers represent numerical results. The dashed line is the proposed fitted formula, \cref{eq:kScPaolo}. \label{fig:giKnl}}
	\end{figure}

	\begin{table*}[t]
		\caption{Analytic values of $ \kappa $ for strongly-coupled \glspl{gimmf} with increasing number of guided modes.}
		\label{tab:kSc}
		\begin{equation*}		
			\begin{array}{*{9}c}
				\toprule
				M & 2 & 6 & 12 & 20 & 30 & 42 & 56 & 72 \\
				\kappa & 8/9 & 40/63 & 56/117 & 8/21 & 88/279 & 104/387 & 40/171 & 136/657\\
				\midrule
				\midrule
				M & 90 & 110 & 132 & 156 & 182 & 210 & 240 & 272\\
				\kappa & 152/819 & 56/333 & 184/1197 & 200/1413 & 8/61 & 232/1899 & 248/2169 & 88/819\\
				\midrule
				\midrule
				M & 306 & 342 & 380 & 420 & 462 & 506 & 552 & 600\\
				\kappa & 280/2763 & 296/3087 & 104/1143 & 328/3789 & 344/4167 & 40/507 & 376/4977 & 392/5409\\
				\midrule
				\midrule
				M & 650 & 702 & 756 & 812 & 870 & 930 & 992 & 1056\\
				\kappa & 136/1953 & 424/6327 & 440/6813 & 152/2439 & 472/7839 & 488/8379 & 56/993 & 520/9513\\
				\bottomrule
			\end{array}
		\end{equation*}
	\end{table*}
	
	Equations \ref{eq:aeffApprR} and \ref{eq:kScPaolo}, with the help of \cref{eq:gamma}, lead to
	\begin{equation}\label{eq:giMyFormula}
		\gamma \kappa \approx \frac{\omega_0 n_2}{c} \frac{M}{M+1} \frac{7}{4} \frac{(\mathrm{NA} k_0)^2}{4\pi M}
	\end{equation}
	whose agreement with the numerical results is visible in Fig.\ref{fig:giCycle}, for the cases $ \Delta = \Delta_\mathrm{max} $ and $ \Delta = \Delta_\mathrm{GISMF} $.
	
	Observe that due to the approximations involved in obtaining \cref{eq:giMyFormula}, the accuracy for $ M = 2 $, i.e., the \gls{smf} case, is quite bad. A better approximation in this case is \cref{eq:giMyFormula2}.

\begin{figure}[!t]
	\centering
	\begin{tikzpicture}
	\begin{axis}[
		enlargelimits=false,
		axis on top,
		axis equal image,
		width = 0.2\textwidth,
		xtick={-40,0,+40},
		ytick={-40,0,+40},
		xlabel={$x [\mu\mathrm{m}]$},
		ylabel={$y [\mu\mathrm{m}]$},
		ylabel style={inner sep=0pt}, 
		tick label style={inner sep=0pt}, 
		]
		\addplot graphics [xmin=-44.000000,xmax=44.000000,ymin=-44.000000,ymax=44.000000] {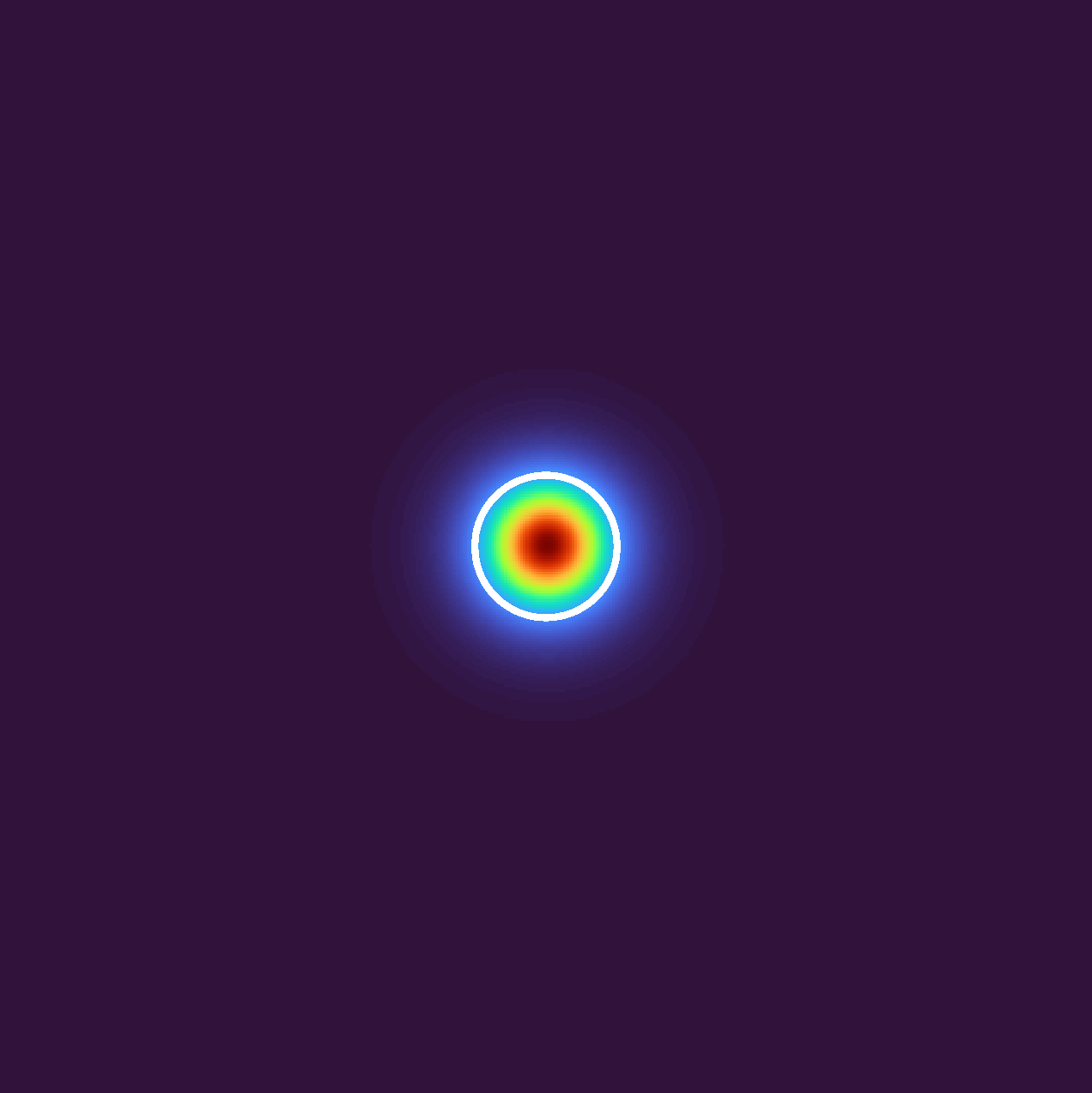};
	\end{axis}
\end{tikzpicture}\hfil
\begin{tikzpicture}
	\begin{axis}[
		enlargelimits=false,
		axis on top,
		axis equal image,
		width = 0.2\textwidth,
		xtick={-40,0,+40},
		ytick={-40,0,+40},
		xlabel={$x [\mu\mathrm{m}]$},
		ylabel={$y [\mu\mathrm{m}]$},
		ylabel style={inner sep=0pt}, 
		tick label style={inner sep=0pt}, 
		]
		\addplot graphics [xmin=-44.000000,xmax=44.000000,ymin=-44.000000,ymax=44.000000] {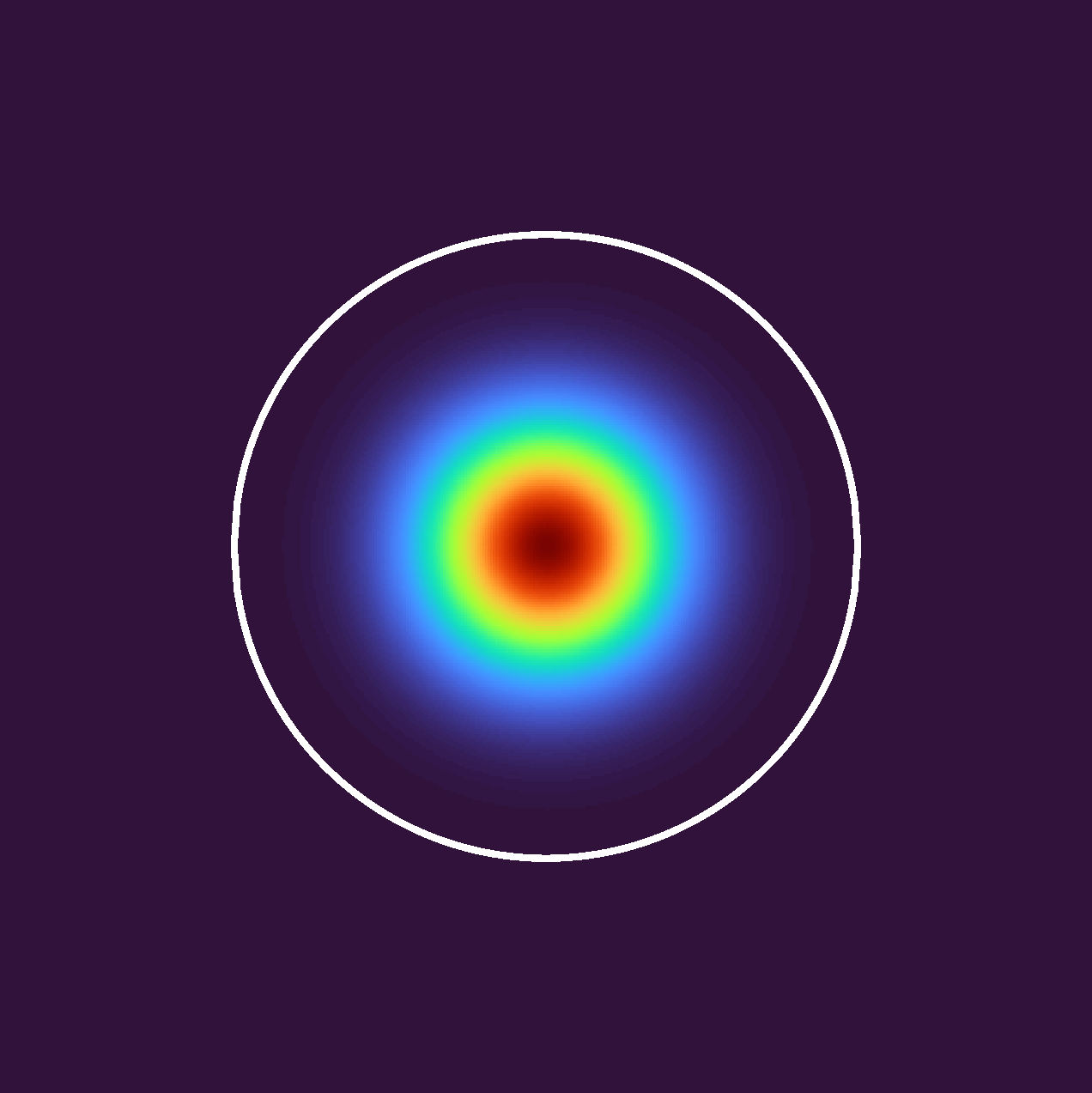};
	\end{axis}
\end{tikzpicture}\hfil
\begin{tikzpicture}
	\begin{axis}[
		enlargelimits=false,
		axis on top,
		axis equal image,
		width = 0.2\textwidth,
	   xtick={-40,0,+40},
		ytick={-40,0,+40},
		xlabel={$x [\mu\mathrm{m}]$},
		ylabel={$y [\mu\mathrm{m}]$},
		ylabel style={inner sep=0pt}, 
		tick label style={inner sep=0pt}, 
		]
		\addplot graphics [xmin=-44.000000,xmax=44.000000,ymin=-44.000000,ymax=44.000000] {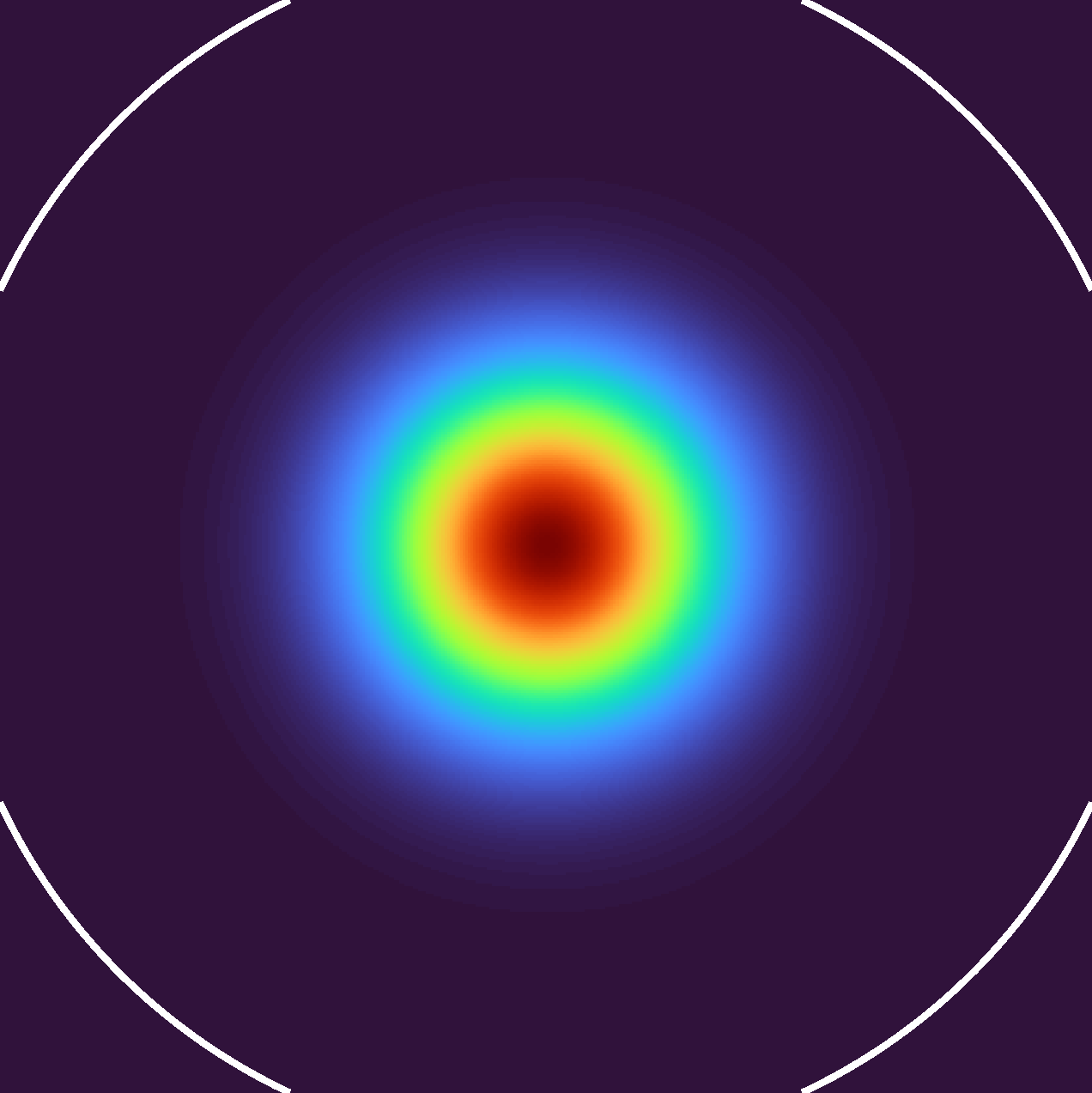};
	\end{axis}
\end{tikzpicture}
	\caption{Intensity profiles of the fundamental modes of a set of fibers with increasing $ R $, fixed $ \Delta $, and, hence, increasing $ \aeff{1}{1} $. The white line indicates the core boundary.}
	\label{fig:modesR}
\end{figure}

	The intuition behind the trend $ \gamma\kappa \propto 1/M $ is that two effects act in the same direction. Firstly, the fundamental mode effective area $ \aeff{1}{1} $, which is a measure of how much a mode spreads over a cross-section of the fiber, increases with $ R $, as exemplified by the modal profiles for three different fibers in \cref{fig:modesR}. Thus, $ \gamma $ reduces as $ 1 / \aeff{1}{1} $. \hlbox{Secondly, by its definition \cref{eq:kSc}, $ \kappa $ is an average of the dimensionless terms $ \aeff{1}{1}/\aeff{a}{b} $, that are the inverses of the mode effective areas $ \aeff{a}{a} $ and of the intermodal effective areas $ \aeff{a}{b} $ (when $ b \ne a $), normalized by the effective area of the fundamental mode. Note that $ \aeff{a}{b} $ is a measure of the overlap between two modes: the larger $ \aeff{a}{b} $, the smaller the overlap. By definition \eqref{eq:aeffInter}, the intemodal area $ \aeff{a}{b} $ is no smaller than $ \max{\{\aeff{a}{a},\aeff{b}{b}\}} $.  We can rewrite $ \kappa $ as $ \kappa = \kappa_\mathrm{intra} + \kappa_\mathrm{inter}$, where the term $ \kappa_\mathrm{intra} \propto 1/N^2 \sum_{a} \aeff{1}{1}/\aeff{a}{a} $ accounts for a (rescaled) average of the inverses of the mode effective areas, and $ \kappa_\mathrm{inter} \propto 1/N^2  \sum_{a}\sum_{b\ne a} \aeff{1}{1}/\aeff{a}{b} $ for a (rescaled) average of inverses of the intermodal areas. Increasing the number of modes brings into play modes with larger $ \aeff{a}{a} $ and $ \aeff{a}{b} $, since the higher order modes are less confined in the core than modes of lower order. Hence, the terms $ \kappa_\mathrm{intra} $ and $ \kappa_\mathrm{inter} $ have to reduce as it can be observed in \cref{fig:areasScaling}. Note that the $ \kappa_\mathrm{intra} $ becomes negligible for $ M \gtrsim 42 $ because the summation runs over $ N $ terms, while the summation in $ \kappa_\mathrm{inter} $ runs over $ N^2-N $ terms. In other words, the contribution of the intermodal effective areas significantly outweighs the contribution of the mode effective areas, even though the average of the mode effective areas (with the $ 1/N $ prefactor) is bigger than the average of the intermodal areas (with the $ 1/(N^2-N) $ prefactor).\footnote{\hlbox{This is in contrast to coupled-core \glspl{mcf} for which the term relative to the intermodal areas vanishes \cite{antonelli16}.}}\footnote{\hlbox{Through fitting, we have observed that the average inverse mode effective area (with the $ 1/N $ prefactor) scales as $ \sim1/M^{0.3} $, and the average inverse intermodal area (with the $ 1/(N^2-N) $ prefactor) as $ \sim1/\sqrt{M} $.}} Since both $ \kappa_\mathrm{intra} $ and $ \kappa_\mathrm{inter} $ reduce with $ M $, so does $ \kappa $.}

\begin{figure}[!t]
	\centering
	\begin{tikzpicture}

\begin{axis}[%
width=0.4\textwidth,
height=0.3\textwidth,
at={(0\textwidth,0\textwidth)},
scale only axis,
xmode=log,
xmin=1,
xmax=3e3,
xminorticks=true,
xlabel={$ M $},
ylabel={$ \kappa,\,\, \kappa_\mathrm{intra},\,\, \kappa_\mathrm{inter} $},
ymode=log,
ymin=5e-05,
ymax=1,
yminorticks=true,
axis background/.style={fill=white},
xmajorgrids,
xminorgrids,
ymajorgrids,
yminorgrids,
legend style={legend cell align=left, align=left, at={(0.03,0.03)},anchor=south west},
]
\addplot [color=mycolor1, only marks, mark=x, mark options={solid, mycolor1}]
  table[row sep=crcr]{%
6	0.148229147576076\\
12	0.0680248115469418\\
20	0.0367584983763565\\
30	0.0222463173447639\\
42	0.0145338348835293\\
56	0.0101239486964385\\
72.0000000000001	0.00734162581661901\\
90.0000000000001	0.00556050726209355\\
110	0.00432265837419147\\
132	0.00345404490057227\\
156	0.00280386594176906\\
2	0.444444444444443\\
20	0.0366958507970347\\
30	0.0222242491711643\\
182	0.00230666251783165\\
210	0.00194050699272031\\
240	0.00162880672438476\\
306	0.00120471115615783\\
420	0.000811439978095772\\
552.000000000001	0.000574206062612841\\
702	0.00042250065543623\\
870.000000000001	0.000321306146711279\\
1190	0.000214271026083962\\
1482	0.000161468609186786\\
1892	0.00011752898395925\\
2352	8.84989404370563e-05\\
42	0.0150264546408451\\
56	0.0106466157774194\\
72.0000000000001	0.00784635603270088\\
90.0000000000001	0.00594600874206465\\
132	0.00368467439530394\\
156	0.00299367854176557\\
182	0.00244161911524607\\
240	0.00172461944147022\\
306	0.00126096689333863\\
380	0.000955896872708847\\
462.000000000001	0.000742529053237424\\
600	0.000530354835271272\\
756	0.000391962263270477\\
992	0.000275079488605091\\
1190	0.000217125669180294\\
1560	0.000151956284301346\\
1892	0.000117860020135495\\
6	0.148453855253799\\
56	0.0101438590333546\\
72.0000000000001	0.00737456077831968\\
90.0000000000001	0.00557907585417426\\
110	0.00434680203944005\\
132	0.0034434696893161\\
};
\addlegendentry{$ \kappa_\mathrm{intra} $}

\addplot [color=mycolor2, only marks, mark=o, mark options={solid, mycolor2}]
  table[row sep=crcr]{%
6	0.440217903002212\\
12	0.390559227771558\\
20	0.327822093745187\\
30	0.284443087151196\\
42	0.246783125533751\\
56	0.219144853895669\\
72	0.195528583959742\\
90	0.177186241230228\\
110	0.1612897320593\\
132	0.148418186970554\\
156	0.137165478769466\\
2	0.444444021328087\\
30	0.283648187921308\\
182	0.127727636594812\\
210	0.119174794768107\\
240	0.111836201181322\\
306	0.099435539884557\\
420	0.0852070434851433\\
552	0.0745744275691378\\
702	0.0662624214797106\\
870	0.0596336972562164\\
1190	0.0510874749316436\\
1482	0.0458326230559023\\
1892	0.0406117190593687\\
2352	0.0364521245128373\\
42	0.246400519229252\\
56	0.218734714674452\\
72	0.195121598796795\\
90	0.176856596436929\\
132	0.148162339834956\\
156	0.136825544637559\\
182	0.127367206873749\\
240	0.11161664450365\\
380	0.0894111619908307\\
462	0.0813050642692091\\
600	0.071535195211708\\
756	0.063885230747589\\
992	0.0558856130678097\\
1560	0.044683948316685\\
6	0.440912306214542\\
132	0.148835052457134\\
};
\addlegendentry{$ \kappa_\mathrm{inter} $}

\addplot [color=mycolor3, only marks, mark=square, mark options={solid, mycolor3}]
  table[row sep=crcr]{%
6	0.588447050578288\\
12	0.458584039318499\\
20	0.364580592121543\\
30	0.30668940449596\\
42	0.26131696041728\\
56	0.229268802592107\\
72	0.202870209776362\\
90	0.182746748492321\\
110	0.165612390433492\\
132	0.151872231871126\\
156	0.139969344711235\\
2	0.888888465772531\\
20	0.363893727264629\\
30	0.305872437092472\\
182	0.130034299112644\\
210	0.121115301760827\\
240	0.113465007905707\\
306	0.100640251040715\\
420	0.0860184834632391\\
552	0.0751486336317507\\
702	0.0666849221351469\\
870	0.0599550034029277\\
1190	0.0513017459577275\\
1482	0.045994091665089\\
1892	0.0407292480433279\\
2352	0.0365406234532744\\
56	0.229381330451871\\
132	0.15184701423026\\
182	0.129808825988995\\
240	0.11334126394512\\
380	0.0903670588635396\\
462	0.0820475933224465\\
600	0.0720655500469792\\
756	0.0642771930108595\\
992	0.0561606925564148\\
1560	0.0448359046009863\\
6	0.589366161468341\\
12	0.458649096522242\\
132	0.15227852214645\\
};
\addlegendentry{$ \kappa $}
\end{axis}
\end{tikzpicture}%
	\caption{\hlbox{Comparison of the contribution of $ \kappa_\mathrm{intra} $, accounting for the mode effective areas, and $ \kappa_\mathrm{inter} $, accounting for the intermodal effective areas, to $ \kappa = \kappa_\mathrm{intra} + \kappa_\mathrm{inter} $. The values have been computed with the numerical approach for the same \glspl{mmf} as in \cref{fig:giKnl}.}\label{fig:areasScaling}}	
\end{figure}
	
\subsubsection{Scaling with the Index Difference}
To study the dependence of $ \gamma \kappa $ on $ \Delta $, we considered the set of \glspl{gimmf} with increasing $ M $, obtained by increasing $ \Delta $ from $ \Delta_\mathrm{GISMF} $ (or from $ \Delta_\mathrm{min,GI} $) to $ \Delta_\mathrm{max} $, fixed $ R $. We recently proposed the following approximate closed-form expressions \cite{carniello23} 
\begin{align}
	\aeff{1}{1} &\approx \frac{\pi R^2}{\sqrt{M}} \label{eq:aeffApprD}\\
	\kappa \approx &\frac{M}{M+1} \frac{7}{4\sqrt{M}} \notag
\end{align}
The derivation of \cref{eq:aeffApprD} follows a similar approach to \cref {eq:aeffApprR} and is given in Appendix \ref{sec:aeffDelta}. The expression for $ \kappa $ is the same as in \cref{sec:scalingR}, since, as proved in Appendix \ref{sec:k}, $ \kappa $ depends on $ M $ and not on $ R $ or on $ \Delta $ alone. Hence, the analytic values for $ \kappa $ are again the ones in \cref{tab:kSc}, displayed in \cref{fig:giKnl}. 

The comparison between the numerical results and the approximate formulas is visible in \cref{fig:giAeff} for $ \aeff{1}{1} $ and in \cref{fig:giKnl} for $ \kappa $, for the cases $ R = R_\mathrm{GISMF} $ and $ R = R_\mathrm{max} $.

Equations \ref{eq:kScPaolo} and \ref{eq:aeffApprD}, with the help of \cref{eq:gamma}, lead to
\begin{equation}\label{eq:giMyFormula2}
	\gamma\kappa \approx \frac{\omega_0 n_2}{c} \frac{M}{M+1} \frac{7}{4} \frac{1}{\pi R^2}
\end{equation}
where we stress that all the parameters (including $ R $) are constant in this scenario (assuming $ M / (M+1) \approx 1 $). The comparison between the numerical results and \cref{eq:giMyFormula2} is given in Fig.\ref{fig:giCycle} for the cases $ R = R_\mathrm{GISMF} $ and $ R = R_\mathrm{max} $, where it can be seen that $ \gamma  \kappa$ remains nearly constant as $\Delta$ is scaled. 

\begin{figure}[!t]
	\centering
	\begin{tikzpicture}
	\begin{axis}[
		enlargelimits=false,
		axis on top,
		axis equal image,
		width = 0.2\textwidth,
		xlabel={$x [\mu\mathrm{m}]$},
		ylabel={$y [\mu\mathrm{m}]$},
		ylabel style={inner sep=0pt}, 
		tick label style={inner sep=0pt}, 
		]
		\addplot graphics [xmin=-55.000000,xmax=55.000000,ymin=-55.000000,ymax=55.000000] {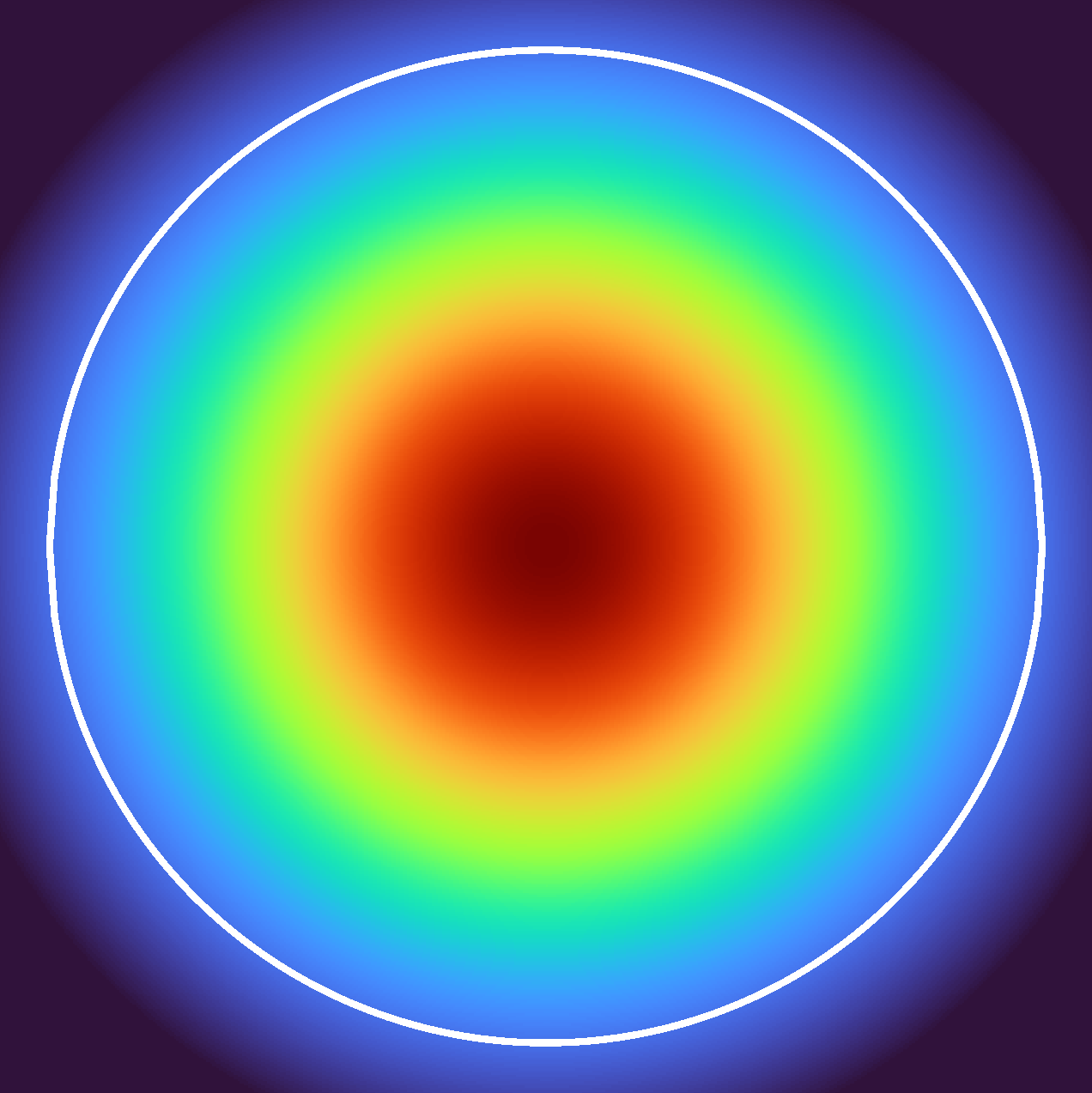};
	\end{axis}
\end{tikzpicture}\hfil
\begin{tikzpicture}
	\begin{axis}[
		enlargelimits=false,
		axis on top,
		axis equal image,
		width = 0.2\textwidth,
		xlabel={$x [\mu\mathrm{m}]$},
		ylabel={$y [\mu\mathrm{m}]$},
		ylabel style={inner sep=0pt}, 
		tick label style={inner sep=0pt}, 
		]
		\addplot graphics [xmin=-55.000000,xmax=55.000000,ymin=-55.000000,ymax=55.000000] {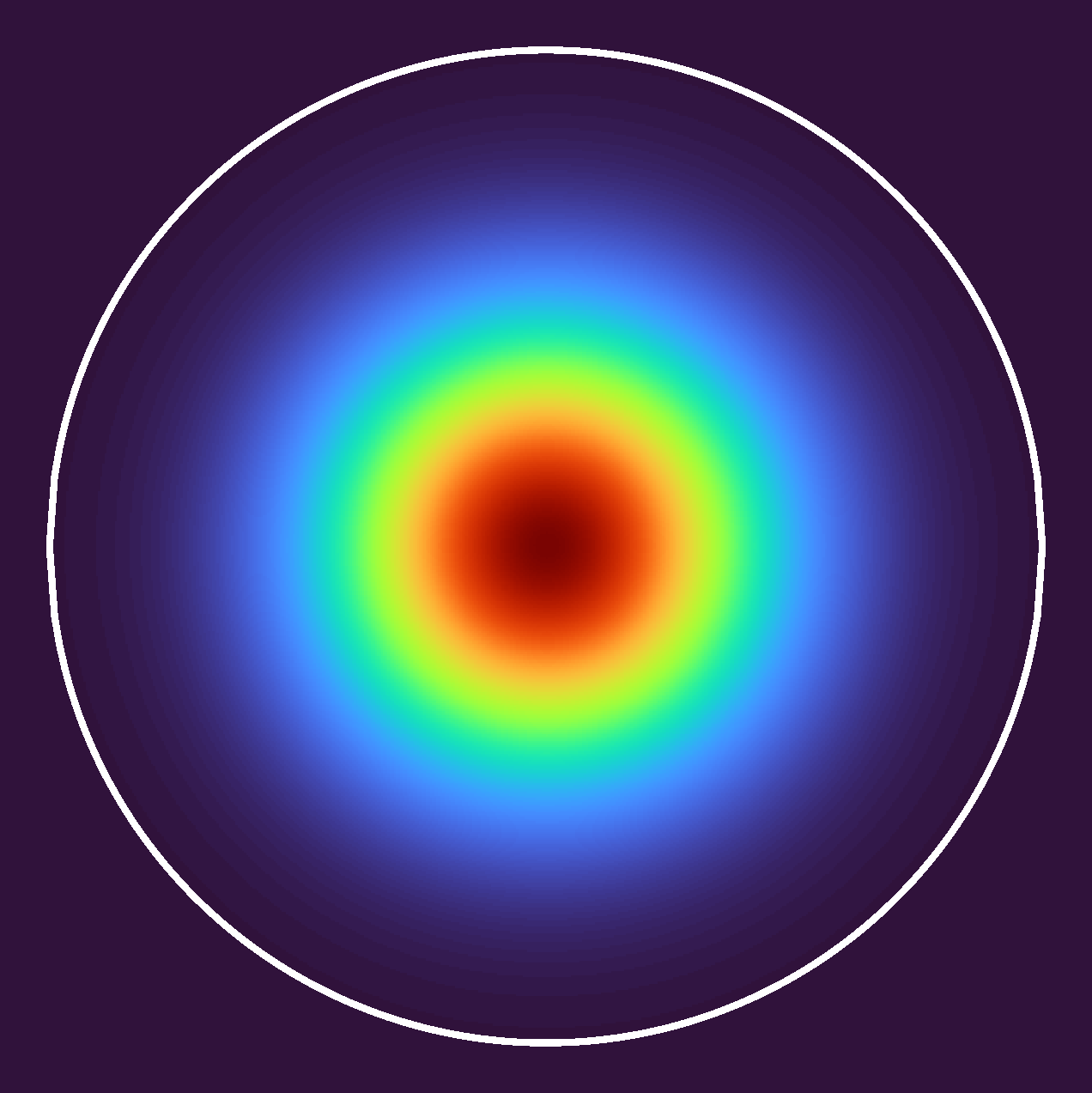};
	\end{axis}
\end{tikzpicture}\hfil
\begin{tikzpicture}
	\begin{axis}[
		enlargelimits=false,
		axis on top,
		axis equal image,
		width = 0.2\textwidth,
		xlabel={$x [\mu\mathrm{m}]$},
		ylabel={$y [\mu\mathrm{m}]$},
		ylabel style={inner sep=0pt}, 
		tick label style={inner sep=0pt}, 
		]
		\addplot graphics [xmin=-55.000000,xmax=55.000000,ymin=-55.000000,ymax=55.000000] {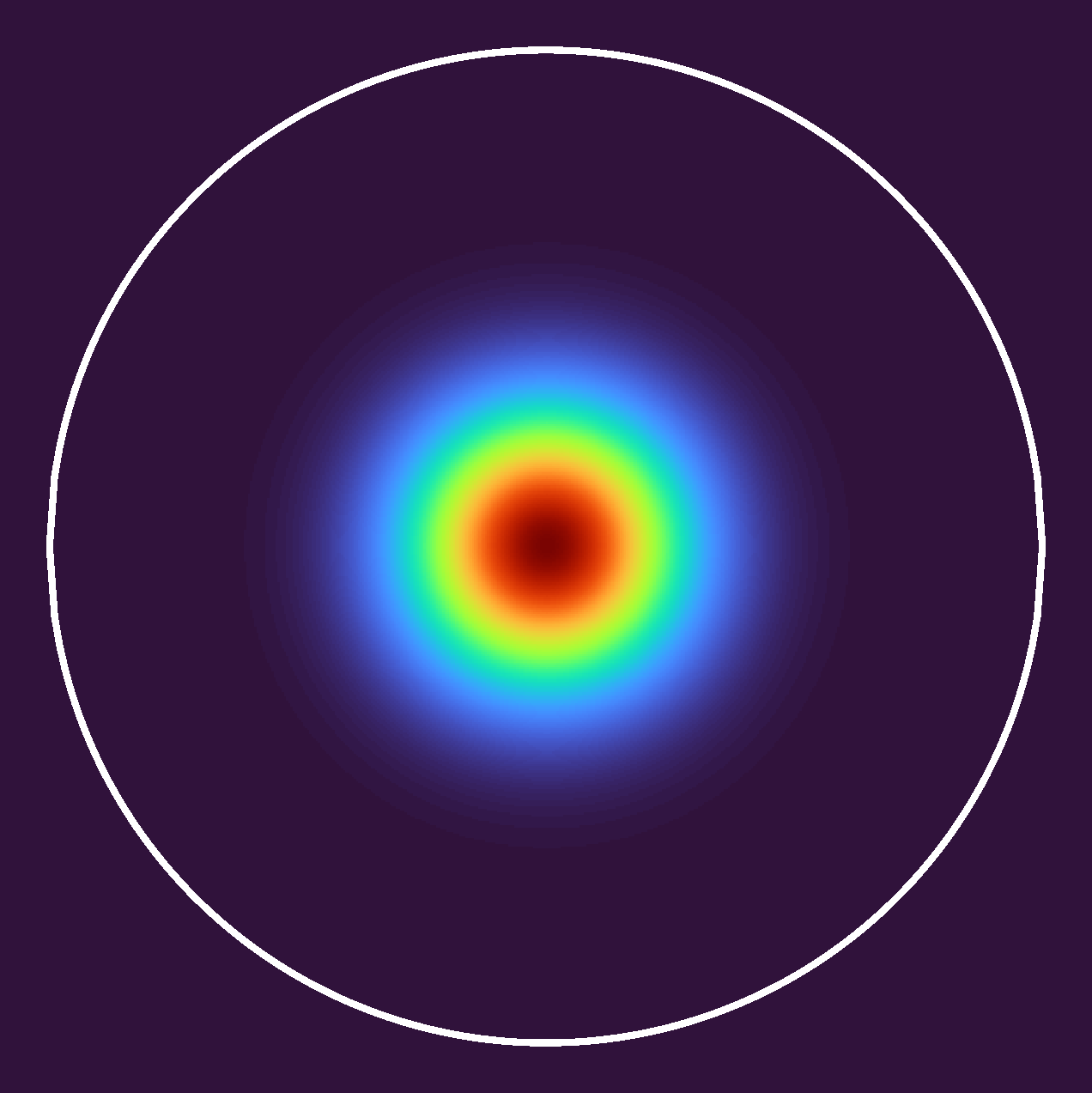};
	\end{axis}
\end{tikzpicture}
	\caption{Intensity profiles of the fundamental modes of a set of fibers with increasing $ \Delta $, fixed $ R $, and, hence, decreasing $ \aeff{1}{1} $. The white line indicates the core boundary.}
	\label{fig:modesDelta}
\end{figure}

The intuition behind the (quasi-)constant behavior of $ \gamma  \kappa$ in this scenario is that two effects balance each other out. The first is that $ \aeff{1}{1} $ reduces with $ \Delta $, since the fundamental mode gets more confined in the core, as exemplified by the modal profiles for three different fibers in Fig.\ref{fig:modesDelta}. Thus, $ \gamma \propto 1/\aeff{1}{1} $ increases. \hlbox{On the other hand, $ \kappa $ decreases with $ M $ for the same qualitative reason as in the previous section. Indeed, given that $ \kappa $ does not depend on the absolute values of the mode effective areas and of the intermodal effective areas, but on their value normalized by the fundamental mode effective area, the reduction of $ \aeff{1}{1} $ and $ \aeff{a}{b} $ does not affect the reduction of the dimensionless $ \aeff{1}{1}/\aeff{a}{b} $ terms. In other words, increasing the number of modes brings still into play modes with smaller $ \aeff{1}{1}/\aeff{a}{b} $ terms. Given that these terms enter the averaging definition of $ \kappa $, $ \kappa $ reduces with $ M $ in the same manner as in \cref{sec:scalingR}.}

It is worth mentioning that  \cref{eq:giMyFormula} and \cref{eq:giMyFormula2} are in fact valid for any \gls{gimmf}, even when $ \Delta $ and $ R $ are varied together. The disadvantage of using the two formulas for a different case than the one they have been developed for is that the link between $ \gamma\kappa $ and $ M $ becomes hidden behind $ \mathrm{NA} $ for \cref{eq:giMyFormula} and behind $ R $ for \cref{eq:giMyFormula2}. Despite that, \cref{eq:giMyFormula2} is a better approximation than \cref{eq:giMyFormula} for small $ M $, also for the cases for which \cref{eq:giMyFormula} has been developed. 

Finally, observe from Fig.\ref{fig:giCycle} that the scaling of $ \gamma \kappa $ with $ M $ is not simply $ 1/M $, but depends on the  strategy to increase $ M $ --  subject to the consideration of the linear effects (e.g., modal dispersion) \cite{sillard14, ferreira24}. The  $ \gamma\kappa \propto 1/M $ scaling is possible only for $ M \lesssim 200 $. A design strategy with poor $ \gamma\kappa $  roll-off with $ M $, as for fixed $ R $, may lead to enhanced nonlinearities, as discussed in \cref{sec:rates}.

\subsection{Weak Coupling Regime}\label{sec:imgcr}
In the case of the \gls{imgcr}, the nonlinear coupling coefficients form a symmetric square matrix $ \gamma \mat{\kappa} $ -- whose size matches the number of mode groups. To find a closed-form expression for that, we can approximate again $ \gamma $ and $ \mat{\kappa} $ separately. The $ \gamma $ coefficient can be expressed through \cref{eq:aeffApprR} or \cref{eq:aeffApprD} as before. In Appendix \ref{sec:k} we derived the analytic $ \kappa_{ab} $ reported in \cref{tab:kImgc} for a \gls{gimmf} up to the first $ 10 $ groups ($ 110 $ polarization modes), regardless of whether $ R $ and/or $ \Delta $ are varied. The analytic $ \kappa_{ab} $ for a \gls{gimmf} up to the first $ 32 $ groups ($ 1056 $ polarization modes) are provided in \cite{mmfData}. In addition, we propose the following simple relation obtained by fitting the numerical results
\begin{equation}\label{eq:kImgcPaolo}
	\kappa_{ab} \approx \frac{4}{3} \frac{M_a}{M_a+\delta_{ab}} \frac{1}{\hlbox{\max{\{a,b\}}}}
\end{equation}
where $ M_a = 2a $ is the number of modes in group $ a $. It can be observed that the fitted formula \eqref{eq:kImgcPaolo} yields the exact same values as the analytic results of \cref{tab:kImgc}. 

\cref{fig:kImgc} compares the analytic/fitted values of $ \mat{\kappa} $ with the numeric ones. \hlbox{The error between them has been computed as}
\begin{equation}\label{eq:errMetric}
	\hlbox{\frac{1}{G^2}\sum_{a=1}^{G} \sum_{b=1}^{G} \frac{\abs{\kappa_{ab} - \tilde{\kappa}_{ab}}}{\kappa_{ab}}}
\end{equation}
\hlbox{where with $ \tilde{\kappa}_{ab} $ we indicated the values obtained with the analytic expression \cref{eq:kImgcPaolo} to distinguish them from the ones obtained numerically, which we indicated as $ \kappa_{ab} $. A plot for the error metric \eqref{eq:errMetric} is reported in \cref{fig:avgErr}. The analytic result is exact for an \gls{smf}, the error is $ \sim7\% $ for a $ 6 $-modes (or $ 2 $ groups) \glspl{gimmf}, and reduces to below $ 0.6\% $ for a $ 992 $-modes (or $ 31 $ groups) \glspl{gimmf}.}

Finally, observe that within our approximations 
the $ \mat{\kappa} $ matrices do not depend on the number of fiber modes, such that for any two \glspl{gimmf} with $ M_1 $ and $ M_2 $ modes respectively, and $ M_2 > M_1 $, $ \mat{\kappa} $ for the $ M_1$-modes fiber is a submatrix of the $ \mat{\kappa} $ for the $ M_2$-modes fiber. Hence, the data in \cref{tab:kImgc} are valid for the $ \kappa_{ab} $ of the first $ 10 $ mode groups of any \gls{gimmf}, and the ones in \cite{mmfData} for the first $ 32 $ mode groups.

\begin{figure}[!t]
	\centering
	\begin{tikzpicture}
		\pgfplotsset{
		layers/my layer set/.define layer set={
			background,
			main,
			foreground
		}{
		},
		set layers=my layer set,
	}
\begin{axis}[
	view={70}{40},
	xtick = {log10(1),log10(10),log10(20),log10(30)},
	xticklabels={1, 10,20,30},
	ytick = {log10(1),log10(10),log10(20),log10(30)},
	yticklabels={1, 10,20,30},
	ztick = {-1,0},
	zticklabels={0.1,1},
	xlabel={$ a $},
	ylabel={$ b $},
	zlabel={$ \kappa_{ab} $},
	grid,
	grid style={gray!80, line width=1pt}, 
	legend style={at={(0.75,0.8)},anchor=south west, opacity = 1}
	]
	\addplot3[%
	surf, fill opacity=0.4, mesh/rows=32,on layer=foreground,
	]%
	file {parts/figures/kImgcAna2.dat};
	\addlegendentry{Analytic};
	
	\addplot3 [%
	color=black, draw=none, mark=*, mark size=1.2, only marks
	]%
	file {parts/figures/kWcrNumData.dat};
	\addlegendentry{Numerical};

\end{axis}

\end{tikzpicture}%
	\caption{Comparison between the analytic $ \mat{\kappa} $ (represented as a surface) and the numerical $ \mat{\kappa} $ for a \gls{gimmf} with $ 31 $ mode groups.\label{fig:kImgc}}
\end{figure}

\begin{table}[t]
	\caption{Analytic values of $ \kappa_{ab} $ for a \gls{gimmf} up to $ 10 $ mode groups. The analytic $ \kappa_{ab} $ for a \gls{gimmf} up to the first $ 32 $ groups ($ 1056 $ modes) are provided in \cite{mmfData}.}
	\label{tab:kImgc}	
	\begin{equation*}
		\begin{array}{|c|cccccccccc|}
			\toprule
			\diagCell{.10em}{.15em}{$ a $}{$ b $} & 1 & 2 & 3 & 4 & 5 & 6 & 7 & 8 & 9 & 10\\
			\midrule
			1 &\frac{8}{9} &\frac{2}{3} &\frac{4}{9} &\frac{1}{3} &\frac{4}{15} &\frac{2}{9} &\frac{4}{21} &\frac{1}{6} &\frac{4}{27} &\frac{2}{15}\\[0.5em] 
			2 &\frac{2}{3} &\frac{8}{15} &\frac{4}{9} &\frac{1}{3} &\frac{4}{15} &\frac{2}{9} &\frac{4}{21} &\frac{1}{6} &\frac{4}{27} &\frac{2}{15}\\[0.5em] 
			3 &\frac{4}{9} &\frac{4}{9} &\frac{8}{21} &\frac{1}{3} &\frac{4}{15} &\frac{2}{9} &\frac{4}{21} &\frac{1}{6} &\frac{4}{27} &\frac{2}{15}\\[0.5em] 
			4 &\frac{1}{3} &\frac{1}{3} &\frac{1}{3} &\frac{8}{27} &\frac{4}{15} &\frac{2}{9} &\frac{4}{21} &\frac{1}{6} &\frac{4}{27} &\frac{2}{15}\\[0.5em] 
			5 &\frac{4}{15} &\frac{4}{15} &\frac{4}{15} &\frac{4}{15} &\frac{8}{33} &\frac{2}{9} &\frac{4}{21} &\frac{1}{6} &\frac{4}{27} &\frac{2}{15}\\[0.5em] 
			6 &\frac{2}{9} &\frac{2}{9} &\frac{2}{9} &\frac{2}{9} &\frac{2}{9} &\frac{8}{39} &\frac{4}{21} &\frac{1}{6} &\frac{4}{27} &\frac{2}{15}\\[0.5em] 
			7 &\frac{4}{21} &\frac{4}{21} &\frac{4}{21} &\frac{4}{21} &\frac{4}{21} &\frac{4}{21} &\frac{8}{45} &\frac{1}{6} &\frac{4}{27} &\frac{2}{15}\\[0.5em] 
			8 &\frac{1}{6} &\frac{1}{6} &\frac{1}{6} &\frac{1}{6} &\frac{1}{6} &\frac{1}{6} &\frac{1}{6} &\frac{8}{51} &\frac{4}{27} &\frac{2}{15}\\[0.5em] 
			9 &\frac{4}{27} &\frac{4}{27} &\frac{4}{27} &\frac{4}{27} &\frac{4}{27} &\frac{4}{27} &\frac{4}{27} &\frac{4}{27} &\frac{8}{57} &\frac{2}{15}\\[0.5em] 
			10 &\frac{2}{15} &\frac{2}{15} &\frac{2}{15} &\frac{2}{15} &\frac{2}{15} &\frac{2}{15} &\frac{2}{15} &\frac{2}{15} &\frac{2}{15} &\frac{8}{63}\\[0.5em] 
			\bottomrule
		\end{array}
	\end{equation*}
\end{table}

\begin{figure}[!t]
	\centering
	\begin{tikzpicture}

\begin{axis}[%
width=0.4\textwidth,
height=0.1\textwidth,
at={(0\textwidth,0\textwidth)},
scale only axis,
xmin=1.8,
xmax=31,
xtick={2,5,10,15,20,25,30},
xminorticks=true,
xlabel style={font=\color{white!15!black}},
xlabel={Number of mode groups},
ymode=log,
ymin=1e-03,
ymax=0.1,
yminorticks=true,
ylabel style={font=\color{white!15!black}},
ylabel={Relative error},
axis background/.style={fill=white},
xmajorgrids,
xminorgrids,
ymajorgrids,
yminorgrids,
legend style={legend cell align=left, align=left, draw=white!15!black}
]
\addplot [color=mycolor1, line width=1.0pt, only marks, mark size=2.0pt, mark=o, mark options={solid, mycolor1}] table[row sep=crcr]{%
x	y\\
1	4.76006129151081e-07\\
2	0.0707848871870499\\
3	0.0389136191022501\\
4	0.0407393252998892\\
5	0.0274800263592267\\
};

\addplot [color=mycolor3, line width=1.0pt, only marks, mark size=2.0pt, mark=x, mark options={solid, mycolor3}] table[row sep=crcr]{%
x	y\\
2	0.0702139222103364\\
3	0.0377026897232436\\
4	0.0383974998024366\\
5	0.0239595755613014\\
6	0.0239143693642961\\
7	0.0169728209035448\\
8	0.0171003384164161\\
9	0.0130358644899824\\
10	0.0129642443460968\\
11	0.0101940625237787\\
12	0.00944899042781171\\
};

\addplot [color=mycolor4, line width=1.0pt, only marks, mark size=2.0pt, mark=star, mark options={solid, mycolor4}] table[row sep=crcr]{%
x	y\\
13	0.00720946538561262\\
14	0.00737328091381978\\
15	0.00656755890267745\\
17	0.0060222452598054\\
20	0.00567389808048531\\
23	0.00493854652701606\\
26	0.00478159954520823\\
29	0.00441835910689704\\
};

\addplot [color=mycolor2, line width=1.0pt, only marks, mark size=2.0pt, mark=triangle, mark options={solid, mycolor2}] table[row sep=crcr]{%
x	y\\
6	0.0247322922651255\\
7	0.0178962973920541\\
8	0.0181679779491084\\
9	0.0143492706099475\\
11	0.0120215453352158\\
12	0.0121639007214829\\
13	0.010420457488874\\
15	0.00922039075440307\\
17	0.00849891108528493\\
19	0.00782790399689753\\
21	0.00725613250388835\\
24	0.00680261501884042\\
27	0.00613379143491363\\
31	0.0056752746243023\\
};

\addplot [color=mycolor4, line width=1.0pt, only marks, mark size=2pt, mark=star, mark options={solid, mycolor4}] table[row sep=crcr]{%
x	y\\
1	1.26220034440238e-10\\
2	0.0687254661325291\\
3	0.0375395942790285\\
4	0.0381887058953089\\
5	0.023738017901596\\
6	0.0236081993569822\\
7	0.016595844000062\\
8	0.0166556918985284\\
9	0.0125619880884206\\
10	0.0125686287343237\\
11	0.00788212718379935\\
};

\end{axis}
\end{tikzpicture}%
	\caption{\hlbox{Average relative error} \eqref{eq:errMetric} between the analytic and the numeric $ \mat{\kappa} $ for each \gls{gimmf} of \cref{fig:giCycle} up to $ 31 $ groups. The \gls{smf} case is not displayed as the error is $ 0 $. Markers as in \cref{fig:giCycle}.\label{fig:avgErr}}
\end{figure}

\subsection{Step-Index Fibers}\label{sec:simmf}
Even though employing \glspl{simmf} for long-haul communications might be even more challenging than utilizing \glspl{gimmf} due to a larger delay spread \cite{puttnam21}, scaling trends are derived here for the sake of completeness. 

The same approach as the one devised to study \glspl{gimmf} is employed here for \glspl{simmf}. The chosen starting point is a fiber with $ R_\mathrm{SISMF} = \qty{4.1}{\mu\meter} $ and $ \Delta_\mathrm{SISMF} = 0.32\% $, to mimic a SSMF with $ \aeff{1}{1} = \qty{85}{\mu\meter} $ \cite{corning}. The other relevant parameters are $ R_\mathrm{max} = \qty{50}{\micro \meter} $, $ \Delta_\mathrm{min,SI} = 0.0034 \% $, and $ \Delta_\mathrm{max} = 5\% $. Based on the numerical results, we found the same trends as \glspl{gimmf} to approximately hold, that is,
\begin{equation}\label{eq:siGammaKnl}
	\gamma \kappa \propto
	\begin{cases}
		1/(M+1), \quad &\text{if $ R $ is varied fixed $ \Delta $}\\
		M/(M+1), \quad & \text{if $ \Delta $ is varied fixed $ R $}
	\end{cases}
\end{equation}
\hlbox{where the proportionality factors can be obtained by fitting the numerical results}, as displayed in \cref{fig:siCycle}. The scaling of $  \aeff{1}{1} $ and $ \kappa $ with number of modes when varying $ R $ and $ \Delta $ are displayed in \cref{fig:siAeff} and in \cref{fig:siKnl}, including the semi-analytic approximation of $ \aeff{1}{1} $ reported in Appendix \ref{sec:simmfExtra}. Given the lower interest in \glspl{simmf} for long-haul communications compared to \glspl{gimmf}, we do not provide an analytic result for $ \kappa $, even though a more quantitative discussion is carried out in Appendix~\ref{sec:simmfExtra}. 

\begin{figure}[!t]
	\centering
	\begin{tikzpicture}

\begin{axis}[%
width=0.38\textwidth,
height=0.3\textwidth,
at={(0\textwidth,0\textwidth)},
scale only axis,
xmode=log,
xmin=1,
xmax=10000,
xminorticks=true,
xlabel style={font=\color{white!15!black}},
xlabel={$ M $},
ymode=log,
ytick={1e-5,1e-4,1e-3},
yticklabels={$10^{-2}$, $10^{-1}$, $10^{0}$},
yminorticks=true,
ylabel style={font=\color{white!15!black}},
ylabel={$ \gamma \kappa  \, (\unit{W^{-1} km^{-1}})$},
axis background/.style={fill=white},
title style={font=\bfseries},
xmajorgrids,
xminorgrids,
ymajorgrids,
yminorgrids,
legend style={legend cell align=left, align=left, draw=white!15!black}
]

\node[] at (axis cs: 2, 8e-4) {SMF};
\node[] at (axis cs: 2.5,	2.5e-5) {LMAFs};

\node[] at (axis cs: 10, 1.5e-3) {$ \textcolor{mycolor1}{\Delta \uparrow} $};
\node[] at (axis cs: 300, 1e-4) {$ \textcolor{mycolor2}{R \uparrow} $};
\node[] at (axis cs: 17, 1e-4) {$ \textcolor{mycolor3}{R \uparrow} $};
\node[] at (axis cs: 20, 2.5e-5) {$ \textcolor{mycolor4}{\Delta \uparrow} $};

\addlegendentry{Fitted}
\addlegendimage{dashed, color = black}

\addplot [color=mycolor1, only marks, mark=o, mark options={solid, mycolor1}]
  table[row sep=crcr]{%
2	0.00109869977884016\\
6	0.00191064973859848\\
12	0.00205217811039663\\
16	0.00202943301864311\\
20	0.00210919605705861\\
24	0.00220231245054091\\
30	0.00212356702020737\\
34	0.00219172487738382\\
};
\addlegendentry{$ R = R_\mathrm{SISMF} $}

\addplot [color=mycolor2, only marks, mark=triangle, mark options={solid, mycolor2}]
  table[row sep=crcr]{%
38	0.00193669296274432\\
54	0.00147786549726302\\
76	0.00101912181738876\\
102	0.000760778392679893\\
144	0.00056457026753803\\
196	0.000416761194070954\\
276	0.000300588702722333\\
382	0.000216188252420503\\
530	0.000158081882262261\\
732	0.000113553914773639\\
1016	8.28640731144314e-05\\
1410	5.94833770092694e-05\\
1956	4.31931057957596e-05\\
2716	3.1016549147539e-05\\
3772	2.24077480573248e-05\\
4770	1.76857181350048e-05\\
};
\addlegendentry{$ \Delta = \Delta_\mathrm{max} $}

\addplot [color=mycolor3, only marks, mark=x, mark options={solid, mycolor3}]
  table[row sep=crcr]{%
6	0.000477081320380875\\
12	0.000285226285410152\\
16	0.000245112993916188\\
20	0.000190674255757759\\
24	0.000164423167082466\\
30	0.000136035990582563\\
34	0.0001181852041814\\
38	0.000109510839940697\\
46	8.94853524364046e-05\\
64	6.90185037514331e-05\\
84	5.19801318767664e-05\\
110	4.06774716056179e-05\\
156	2.90838579821223e-05\\
210	2.20226382502083e-05\\
280	1.69417017666108e-05\\
};
\addlegendentry{$ \Delta = \Delta_\mathrm{SISMF} $}

\addplot [color=mycolor4, only marks, mark size=2.0pt, mark=star, mark options={solid, mycolor4}]
  table[row sep=crcr]{%
284	1.688215580990224e-05\\
354	1.70129151118639e-05\\
440	1.70610285529104e-05\\
542.000000000001	1.71829490439685e-05\\
676.000000000001	1.72594773524831e-05\\
834	1.73202247705161e-05\\
1038	1.73880018747395e-05\\
1284	1.74489500079506e-05\\
1596	1.74863349553143e-05\\
1976	1.75351671250503e-05\\
2452	1.75775713799874e-05\\
3042	1.76146757002352e-05\\
3772	1.7649997952126e-05\\
};
\addlegendentry{$ R = R_\mathrm{max} $}

\addplot [color=mycolor4, only marks, mark size=2.0pt, mark=star, mark options={solid, mycolor4}, forget plot]
  table[row sep=crcr]{%
2	1.01744781812908e-05\\
6	1.28647266664811e-05\\
12	1.38545261840247e-05\\
16	1.37331678231304e-05\\
20	1.42699315695442e-05\\
24	1.48980935772126e-05\\
30	1.44069296223778e-05\\
34	1.5406963867778e-05\\
64	1.56029568033247e-05\\
110	1.63767090910377e-05\\
234	1.68035783212169e-05\\
440	1.70610285529104e-05\\
834	1.73202247705161e-05\\
1584	1.74748065315819e-05\\
3002	1.76164190509964e-05\\
};

\addplot [color=mycolor1, dashed, forget plot]
  table[row sep=crcr]{%
2	0.00141571134680492\\
6	0.00182020030303489\\
12	0.00196021571096065\\
16	0.00199865131313635\\
20	0.00202244478114988\\
24	0.00203862433939908\\
30	0.0020550648582652\\
34	0.00206289367677288\\
};
\addplot [color=mycolor2, dashed, forget plot]
  table[row sep=crcr]{%
38	0.00209449215481813\\
54	0.00148518534614376\\
76	0.00106084667581697\\
102	0.000793060136290359\\
144	0.000563346165778669\\
196	0.000414645655014756\\
276	0.000294892397248762\\
382	0.000213277269028478\\
530	0.000153832757133535\\
732	0.000111439555304102\\
1016	8.03197581493677e-05\\
1410	5.78917037830666e-05\\
1956	4.17400071731768e-05\\
2716	3.00644806911693e-05\\
3772	2.16499321595301e-05\\
4770	1.71211892764425e-05\\
};
\addplot [color=mycolor3, dashed, forget plot]
  table[row sep=crcr]{%
6	0.000583011388210984\\
12	0.000313929209036684\\
16	0.000240063512792758\\
20	0.000194337129403661\\
24	0.000163243188699075\\
30	0.000131647732821835\\
34	0.000116602277642197\\
38	0.000104643069678895\\
46	8.6831483350572e-05\\
64	6.27858418073367e-05\\
84	4.80127025585516e-05\\
110	3.67664839412332e-05\\
156	2.59941383278783e-05\\
210	1.93416100354355e-05\\
280	1.45234153646864e-05\\
};
\addplot [color=mycolor4, dashed, forget plot]
  table[row sep=crcr]{%
2	1.09178060606918e-05\\
6	1.40371792208895e-05\\
12	1.5116962237881e-05\\
16	1.54133732621532e-05\\
20	1.55968658009883e-05\\
24	1.57216407273962e-05\\
30	1.58484281526172e-05\\
34	1.59088031170081e-05\\
64	1.61247597204064e-05\\
110	1.62291711712987e-05\\
234	1.63070209672461e-05\\
440	1.63395736962735e-05\\
834	1.6357096265779e-05\\
1584	1.6366376782463e-05\\
4770	1.6371255641457e-05\\
};
\end{axis}
\end{tikzpicture}%
	\caption{Scaling of $ \gamma \kappa $ with $ M $ for different \gls{simmf} designs. Markers indicate numerical results. The dashed lines are the trends in \cref{eq:siGammaKnl} with fitted proportionality factors. \label{fig:siCycle}}
\end{figure}

\begin{figure}[!t]
	\centering
	\pgfplotsset{
	/pgfplots/markString/.style={
		legend image code/.code={
			\node[color=mycolor1] (n1) {\pgfuseplotmark{o}};
			\node[color=mycolor2] (n2) [right=0.1cm of n1] {\pgfuseplotmark{triangle}};
			\node[color=mycolor3] (n3) [right=0.1cm of n2] {\pgfuseplotmark{x}};
			\node[color=mycolor4] (n4) [right=0.1cm of n3] {\pgfuseplotmark{star}};
			\node[color=mycolor5] (n5) [right=0.1cm of n4] {\pgfuseplotmark{square}};
		}
	}
}
\begin{tikzpicture}

\begin{axis}[%
width=0.37\textwidth,
height=0.3\textwidth,
at={(0\textwidth,0\textwidth)},
scale only axis,
xmode=log,
xmin=1,
xmax=10000,
xminorticks=true,
xlabel style={font=\color{white!15!black}},
xlabel={$ M $},
ymode=log,
ymin=1e-11,
ymax=1.45e-08,
ytick={1e-11,1e-10,1e-09, 1e-8},
yticklabels={$10^{1}$, $10^{2}$, $10^{3}$, $10^{4}$},
yminorticks=true,
ylabel style={font=\color{white!15!black}},
ylabel={$ \aeff{1}{1}  \, (\unit{um^2})$},
axis background/.style={fill=white},
title style={font=\bfseries},
xmajorgrids,
xminorgrids,
ymajorgrids,
yminorgrids,
legend style={legend cell align=left, align=left, draw=white!15!black, at = {(0.99, 0.03)}, anchor = south east}
]

\addlegendentry{Semi-Analytic}
\addlegendimage{dashed}

\node[] at (axis cs: 2, 5e-11) {SMF};
\node[] at (axis cs: 2.8,	3e-9) {LMAFs};
\node[] at (axis cs: 10, 2e-11) {$ \textcolor{mycolor1}{\Delta \uparrow} $};
\node[] at (axis cs: 2000, 7e-10) {$ \textcolor{mycolor2}{R \uparrow} $};
\node[] at (axis cs: 50, 4e-10) {$ \textcolor{mycolor3}{R \uparrow} $};
\node[] at (axis cs: 30, 7.5e-9) {$ \textcolor{mycolor4}{\Delta \uparrow} $};

\addplot [color=mycolor1, only marks, mark=o, mark options={solid, mycolor1}]
  table[row sep=crcr]{%
2	8.52686916171379e-11\\
6	4.24629408518965e-11\\
12	3.67900323605819e-11\\
16	3.577746484465e-11\\
20	3.40001723030566e-11\\
24	3.30148358161537e-11\\
30	3.22941485141898e-11\\
34	3.18493956602594e-11\\
};
\addlegendentry{$ R = R_\mathrm{SISMF} $}

\addplot [color=mycolor2, only marks, mark=triangle, mark options={solid, mycolor2}]
  table[row sep=crcr]{%
38	3.69931355437469e-11\\
54	4.79825155956062e-11\\
76	6.853220637194e-11\\
102	9.20524713034883e-11\\
144	1.22140861202254e-10\\
196	1.66151362637514e-10\\
276	2.28744163132075e-10\\
382	3.17273493262521e-10\\
530	4.32259136943801e-10\\
732	6.01200156087673e-10\\
1016	8.20778412946656e-10\\
1410	1.14150551697314e-09\\
1956	1.56847103869389e-09\\
2716	2.18086768085046e-09\\
3772	3.01448815523781e-09\\
4770	3.81650691200261e-09\\
};
\addlegendentry{$ \Delta = \Delta_\mathrm{max} $}

\addplot [color=mycolor3, only marks, mark=x, mark options={solid, mycolor3}]
  table[row sep=crcr]{%
6	1.70377055619423e-10\\
12	2.66318056417496e-10\\
16	2.98765773348713e-10\\
20	3.79988066551426e-10\\
24	4.47226378800155e-10\\
30	5.11665384224963e-10\\
34	6.14199554991198e-10\\
38	6.44390767646129e-10\\
46	7.99326938810447e-10\\
64	1.0077911495486e-09\\
84	1.35353759650631e-09\\
110	1.71842255498231e-09\\
156	2.38630346085507e-09\\
210	3.13035388129485e-09\\
280	4.05758125131265e-09\\
};
\addlegendentry{$ \Delta = \Delta_\mathrm{SISMF} $}

\addplot [color=mycolor4, only marks, mark size=2.0pt, mark=star, mark options={solid, mycolor4}]
  table[row sep=crcr]{%
284	4.069208191632082e-09\\
354.000000000001	4.02928584257628e-09\\
440	4.00084377665595e-09\\
542	3.97228150785108e-09\\
676.000000000001	3.94787987002857e-09\\
834	3.92669958524417e-09\\
1038	3.90642057503183e-09\\
1284	3.889225108386e-09\\
1596	3.87403788187138e-09\\
1976	3.86009522958893e-09\\
2452	3.84757634405683e-09\\
3042.00000000001	3.83642683163677e-09\\
3772	3.82637626489799e-09\\
};
\addlegendentry{$ R = R_\mathrm{max} $}

\addplot [color=mycolor4, only marks, mark size=2.0pt, mark=star, mark options={solid, mycolor4}, forget plot]
  table[row sep=crcr]{%
2	9.20782063696652e-09\\
6	6.32512924837966e-09\\
12	5.48865349323348e-09\\
16	5.33994475438033e-09\\
20	5.08291932487917e-09\\
24	4.94022512080129e-09\\
30	4.83680209988021e-09\\
34	4.71489046205423e-09\\
64	4.45920999982121e-09\\
110	4.26789878368023e-09\\
234	4.1028430237723e-09\\
440	4.00084377665595e-09\\
834	3.92669958524417e-09\\
1584	3.87487993947963e-09\\
3002	3.83691755035493e-09\\
};

\addplot [color=mycolor3, dashed]
  table[row sep=crcr]{%
6	1.65403640299307e-10\\
12	2.60286401415072e-10\\
16	3.22341920267383e-10\\
20	3.83762134693287e-10\\
24	4.44714825494989e-10\\
30	5.3549526133577e-10\\
34	5.95682897793908e-10\\
38	6.55659376102867e-10\\
46	7.75106201476264e-10\\
64	1.04211288127472e-09\\
84	1.33694430220035e-09\\
110	1.71842255498231e-09\\
156	2.39031885755564e-09\\
210	3.17612395969669e-09\\
280	4.19196546450381e-09\\
};

\addplot [color=mycolor2, dashed]
  table[row sep=crcr]{%
38	3.66473670009055e-11\\
54	4.99705756399795e-11\\
76	6.81453549512438e-11\\
102	8.94981084708417e-11\\
144	1.2382291783694e-10\\
196	1.66151362637514e-10\\
276	2.31064281080659e-10\\
382	3.1685170852312e-10\\
530	4.36390610484567e-10\\
732	5.99287768481921e-10\\
1016	8.28025905108551e-10\\
1410	1.14504607873087e-09\\
1956	1.58402745904541e-09\\
2716	2.19469115863979e-09\\
3772	3.04278587387425e-09\\
4770	3.84403963035382e-09\\
};
\addplot [color=mycolor1, dashed]
  table[row sep=crcr]{%
2	9.22218163261896e-11\\
6	5.01139683836515e-11\\
12	3.94307660569185e-11\\
16	3.66236637199473e-11\\
20	3.48816569878285e-11\\
24	3.36849066082652e-11\\
30	3.24488536620054e-11\\
34	3.18493956602594e-11\\
};
\addplot [color=mycolor4, dashed]
  table[row sep=crcr]{%
2	1.34715763940928e-08\\
6	7.32054713717225e-09\\
12	5.75996654993801e-09\\
16	5.34991071841124e-09\\
20	5.09544189849822e-09\\
24	4.92062302368965e-09\\
30	4.74006290943441e-09\\
34	4.65249529704854e-09\\
64	4.32398838219785e-09\\
110	4.14846096029551e-09\\
234	3.99996125280016e-09\\
440	3.92739297220693e-09\\
834	3.88209387481559e-09\\
1584	3.8541154254092e-09\\
4770	3.83691755035493e-09\\
};
\end{axis}
\end{tikzpicture}%
	\caption{Scaling of $ \aeff{1}{1} $ with $ M $ for different \gls{simmf} designs. Markers indicate numerical results. \hlbox{The dashed lines are theoretical trends (but with fitted proportionality factors!): the line relative to the circles is \cref{eq:siAeffScaling2} with $ R = R_\mathrm{SISMF} $; the line relative to the triangles is \cref{eq:siAeffScaling} with $ \Delta = \Delta_\mathrm{max} $; the line relative to the crosses is \cref{eq:siAeffScaling} with $ \Delta = \Delta_\mathrm{SISMF} $; the line relative to the stars is \cref{eq:siAeffScaling2} with $ R = R_\mathrm{max} $.}\label{fig:siAeff}}
\end{figure}

\begin{figure}[!t]
	\centering
	\begin{tikzpicture}

\begin{axis}[%
width=0.4\textwidth,
height=0.3\textwidth,
at={(0\textwidth,0\textwidth)},
scale only axis,
xmode=log,
xmin=1,
xmax=10000,
xminorticks=true,
xlabel style={font=\color{white!15!black}},
xlabel={$ M $},
ymin=0.6,
ymax=0.95,
yminorticks=true,
ylabel style={font=\color{white!15!black}},
ylabel={$ \kappa $},
axis background/.style={fill=white},
title style={font=\bfseries},
xmajorgrids,
xminorgrids,
ymajorgrids,
yminorgrids,
legend style={legend cell align=left, align=left, draw=white!15!black}
]


\node[] at (axis cs: 2, 0.91) {SMF};

\addplot [color=mycolor1, only marks, mark=o, mark options={solid, mycolor1}]
  table[row sep=crcr]{%
2	0.888888153089679\\
6	0.769785329060621\\
12	0.716347422464657\\
16	0.688910441981669\\
20	0.680419185218127\\
24	0.689868885363126\\
30	0.650681249516726\\
34	0.662316714235957\\
};
\addlegendentry{$ R = R_\mathrm{SISMF} $}

\addplot [color=mycolor2, only marks, mark=triangle, mark options={solid, mycolor2}]
  table[row sep=crcr]{%
38	0.679767504923091\\
54	0.672816146134594\\
76	0.662673028848921\\
102	0.664464990482728\\
144	0.654270685832924\\
196	0.65700649493702\\
276	0.652380778887029\\
382	0.650795232870696\\
530	0.648342904430093\\
732	0.647738449898468\\
1016	0.64531340149506\\
1410	0.64424656714855\\
1956	0.64279045463086\\
2716	0.641802309629546\\
3772	0.640900007125854\\
4770	0.640423462472347\\
};
\addlegendentry{$ \Delta = \Delta_\mathrm{max} $}

\addplot [color=mycolor3, only marks, mark=x, mark options={solid, mycolor3}]
  table[row sep=crcr]{%
6	0.77122660512282\\
12	0.720723429627109\\
16	0.694825357666788\\
20	0.687449023916045\\
24	0.697700288796892\\
30	0.660417148674995\\
34	0.688733312142415\\
38	0.669552909170011\\
46	0.678664112525674\\
64	0.659955362206465\\
84	0.667553661427818\\
110	0.663227303493859\\
156	0.658500589732112\\
210	0.654095653817641\\
280	0.652232956471569\\
};
\addlegendentry{$ \Delta = \Delta_\mathrm{SISMF} $}

\addplot [color=mycolor4, only marks, mark=star, mark options={solid, mycolor4}]
  table[row sep=crcr]{%
284	0.651802911574941\\
354	0.650407131875673\\
440	0.647642417431781\\
542	0.647613961042675\\
676	0.646502265770144\\
834	0.645297053542627\\
1038	0.644476608171399\\
1284	0.643888785401086\\
1596	0.642748598142671\\
1976	0.642223821158683\\
2452	0.641689005030199\\
3042	0.641180127434746\\
3772	0.640782754676692\\
};
\addlegendentry{$ R = R_\mathrm{max} $}

\addplot [color=mycolor4, only marks, mark=star, mark options={solid, mycolor4}, forget plot]
  table[row sep=crcr]{%
2	0.888888888855567\\
6	0.7720553726104\\
12	0.721499398914834\\
16	0.695802481201125\\
20	0.688198289805828\\
24	0.698322313647346\\
30	0.661162533060895\\
34	0.689234730529434\\
64	0.660151042235659\\
110	0.663161381464686\\
234	0.654131641801288\\
440	0.647642417431782\\
834	0.645297053542625\\
1584	0.642464460827383\\
3002	0.641325607684305\\
};

\end{axis}
\end{tikzpicture}%
	\caption{Scaling of $ \kappa $ with $ M $ for different \gls{simmf} designs. \label{fig:siKnl}}
\end{figure}
\section{Optimized and Manufactured Fibers}\label{sec:comparison}
In \cref{fig:giCycleWithFilipe} the $ \gamma\kappa $ values for different optimized and/or manufactured \glspl{gimmf} described in the literature have been reported against the developed framework, assuming to operate in the \gls{scr}. Such fibers have been emulated based on the available information in the respective references, which did not always include all the necessary data about the refractive index profile. Hence, some little discrepancies between the actual $ \gamma\kappa $ and the one computed by us are possible.
Nevertheless, the consistency between the $ \gamma\kappa $ computed for the fibers in the literature and our framework supports the validity of our investigation. In particular, the optimized fibers lie within the foreseen boundaries, and the approximate trends for $ \gamma\kappa $ ($ \propto 1/M $ with $ R $, and $ \mathrm{const} $ with $ \Delta $), are verified again. Similar considerations hold for $ \aeff{1}{1} $ and $ \kappa $.

It should be noted that the optimized and manufactured fibers in general exploit more advanced index profiles, which would formally require an extension to the analysis in this paper for a proper comparison. At the same time, we verified that the $ \kappa $ computed (through the numerical method) for the literature fibers, which exploit trenches and grading indices slightly different from $ 2 $, are close to the $ \kappa $ of the trenchless parabolic \glspl{gimmf}. In addition, a quantitative reasoning based on the Gaussian approximation for the effective area of a generic \gls{gi} profile suggests that for realistic slightly non-parabolic \glspl{gi} $ \gamma\kappa $ deviates by at most $ 10\% $ from the parabolic case, see Appendix \ref{sec:realFibersApp}.

\begin{figure*}[t]
	\centering	
	\input{parts/figures/gkSc_litMmfs.tex}
	\caption{Scaling of $ \gamma\kappa $ with $ M $ for a number of optimized and manufactured \glspl{gimmf}. \hlbox{The notation $ R \!\in\! [a\!\!:\!\!b\!\!:\!\!c] $ in the legend indicates that, moving from top to bottom along a line, the first point refers to a fiber with $ R = a $, the second point refers to a fiber with $ R = a+b $, and so on until the last fiber with $ R = c $. Dashed lines and crosses refer to the same lines and points as in \cref{fig:giCycle}.}\label{fig:giCycleWithFilipe}}
\end{figure*}

The fiber data used for the plots of the paper, including the fibers emulated from the literature, and additional data like the intermodal effective areas $ \aeff{a}{b} $ have been provided in \cite{mmfData}.
\section{Considerations on Data Rates}\label{sec:rates}
The ultimate throughput limit of an \gls{sdm} system in the \gls{scr} depends on Kerr nonlinearity, which depends also on $ \gamma \kappa $. In order to study the impact of the scaling of $ \gamma \kappa $ with $ M $ on the data rates of \gls{gimmf} systems, let us consider the perturbative model developed in \cite{garcia22}. The model assumes a strongly-coupled \gls{sdm} fiber with only strong linear coupling, mode-independent \gls{cd} and Kerr nonlinearity as distortions. No other linear effects like modal delay or \gls{mdl} are included. Ideal distributed amplification able to perfectly compensate the fiber loss is assumed. Circularly-symmetric complex Gaussian modulation is considered. Compensation of the linear effects (coupling and \gls{cd}) is assumed through, e.g., \gls{dsp}. Digital back-propagation (DBP) is potentially applied on the channel under test, so that \gls{spm} is removed. The mean phase noise is also removed. In such case, the achievable information rate in terms of $ \unit{\bit/\s/\Hz/mode} $ for an \gls{awgn} receiver can be computed as 
\begin{equation}\label{eq:rates}
	R(M, P) = \log_2 \left( 1 +  \frac{P}{N_\mathrm{ASE}B +  \sigma_\mathrm{NLI}^2} \right)
\end{equation}
where $ P $ is the transmit power per channel per mode, $ N_\mathrm{ASE} $ is the \gls{ase} noise flat \gls{psd} over a band $ B = 1/T $, $ T $ is the symbol time, and $\sigma_\mathrm{NLI}^2 $ is the variance of the equivalent nonlinear interference noise. It holds $\sigma_\mathrm{NLI}^2 = (\gamma \kappa)^2\left(\eta_2 M^2 + \eta_1 M  + \eta_0\right) \! P^3$, where $ \eta_x $ for $ x \in \{0,1,2\} $ are coefficients whose computation is described in \cite{garcia22}. For the examples below we considered a root-raised cosine shaping pulse with roll-off factor $ 0.1 $ and no DBP.

Neglecting $ \eta_2 $ (realistic for $ M \le 1000 $) and $\eta_0$ (rendered negligible for $M > 10$ since $\eta_0 = \eta_1 $ in our setup \cite{garcia22}), and fixing the power per mode $P$, it follows that the scaling of the nonlinear interference $\sigma_\mathrm{NLI}^2$, and thus of the data rate per mode, depends on $(\gamma \kappa)^2\,\left(\eta_1 M\right)$. In particular, if $ \gamma \kappa $ reduces with number of modes faster than $1/\sqrt{M}$, the data rate increases with the number of modes. This would imply that a \gls{mmf} could achieve a higher data rate than a bundle of \glspl{smf} with same total number of modes, fixed the same power per mode. This is depicted in \cref{fig:rates} for a) the best-case scenario $ \gamma \kappa \propto 1/M $ and for b) the (approximate) limiting one $ \gamma \kappa \propto 1/\sqrt{M} $; $\eta_0$ and $\eta_2$ were not neglected in \cref{fig:rates}, and the parameter setup of \cite[Table I-II] {garcia22} was used. The different data rate scaling trends in \cref{fig:rates1} and \cref{fig:rates2} highlight the importance of designing a fiber keeping into account nonlinearity and, hence, $ \gamma \kappa $. \hlbox{Previously \cite{antonelli16, antonelli16ecoc}, the data rate of \glspl{mmf} was believed to increase with $ M $, while as we showed this is valid only when $ \gamma \kappa \propto 1/M $, which is correct only for a specific way of designing fibers, that is, by only increasing $ R $.} Note that similar conclusions could be reached with other perturbative models in the literature like the ones in \cite{antonelli16, shtaif22, serena22}.

\hlbox{Observe that \eqref{eq:rates} can be exploited to assess once more the accuracy of the derived analytic and fitted expressions for $ \gamma \kappa $ in the \gls{scr}. We verified that using \cref{eq:gamma} for computing $ \gamma $ and the analytic values of \cref{tab:kSc} for computing $ \kappa $, instead of the numerical $ \gamma \kappa $, would introduce an error in the rate computation below $ 0.8 \, \% $ for a $ 6 $-modes \gls{mmf} and below $ 0.02 \, \% $ for a $ 992 $-modes \gls{mmf}. Using \cref{eq:gamma} for computing $ \gamma $ and the fitted values \eqref{eq:giMyFormula2} for $ \kappa  $, instead of the numeric $ \gamma \kappa $, would introduce an error below $ 2.3 \, \% $ for a $ 6 $-modes \gls{mmf} and below $ \sim 0.2 \, \% $ for a $ 992 $-modes \gls{mmf}.
The error metric we adopted is $ \abs{R-\tilde{R}}/R $, where in this case by $ R $ we mean the rate computed via the numeric $ \gamma \kappa $, while by $ \tilde{R} $ we mean the rate computed via either the analytic or the fitted $ \gamma \kappa $. The low error confirms the validity of the analytic and fitted formulas for the described scenario, in particular for \glspl{mmf} with a not too small $ M $.}

It is worth mentioning that \gls{sdm} systems are being investigated in the literature for very different transmission scenarios which might not correspond to the one presented in this section. In particular, the data rate per mode with fixed the power per mode might not be the relevant metric in all situations. Thus, the example discussed above is meant as an application of the $ \gamma \kappa $ study rather than a general statement on the usefulness of \gls{sdm} systems.

\hlbox{Finally, data rates have been computed in this Section considering only chromatic-dispersion and Kerr nonlinear effect, since the purpose is to highlight the role of $ \gamma\kappa $ and of fiber design in such scenarios, rather then proposing accurate estimates of data rates for \gls{sdm} systems. However, more relevant effects which reduce data rates and which we neglected are the presence of potential uncompensated inter-group linear coupling in the \gls{imgcr} \cite{lasagni24ofc}, and in general any residual presence of linear coupling \cite{barbosa24}, which should be accounted for in a thorough rate analysis of \gls{sdm} systems. Other impairments, which have been neglected, but which are expected to play a non-negligible role in terms of rate reduction are \gls{mdl}, bend losses, and coating losses \cite{ferreira24}. The interplay between the mention linear effects (with the inclusion of mode delay) and Kerr nonlinear effects should also be considered, as already partially done in, e.g., \cite{rademacher16, antonelli16, serena22, lasagni23}. Even though complete compensation of the linear effects has been assumed for \cref{eq:rates}, the computational burden of the \gls{dsp} algorithms for increasing number of modes has not been addressed, which is a well-known open challenge \cite{puttnam21, sillard22}.}

\begin{keysubfigs}*{2}
	{
		c = {Comparison between the achievable data rates computed with \cref{eq:rates} when $ \gamma \kappa \propto 1/M$ (\ref{fig:rates1}), and when $ \gamma \kappa \propto 1/\sqrt{M} $ (\ref{fig:rates2}), within the perturbation model and the assumptions of \cite{garcia22}.},
		l = fig:rates
	}		
	\keyfigbox{
		l = fig:rates1,
		lw = 1,
	}{\input{parts/figures/rates1}}
	\keyfigbox{
		l = fig:rates2,
		lw = 1,
	}{\input{parts/figures/rates2}}
\end{keysubfigs}
\section{Side Remarks}\label{sec:sideRemarks}
\hlbox{Some aspects related to assumptions made throughout the papers are worth to be discussed.}

\hlbox{Firstly, we assumed the hypotheses of Manakov equations for the \gls{scr} and the \gls{imgcr} to be fulfilled, i.e., the right levels of \gls{xt} (a metric for the intensity of linear coupling \cite{ferreira17}) among the various mode groups, and a not too large difference between the maximum and the minimum mode delays for strongly coupled modes \cite{antonelli16, ferreira19}. The level of \gls{xt} between two modes depends on both the phase-mismatch between them and the fiber perturbations inducing linear coupling \cite{marcuse74, ho12, palmieri14, ferreira17}. Hence, given a certain fiber, there is no a-priori regime of linear coupling in which it operates; the regime depends on the level of \gls{xt}, which depends on the environment and on the system architecture. In practical scenarios the two regimes which we considered might be hard to achieve, with the intermediate coupling regime being often the actual one \cite{ferreira17,ferreira19, buch19}. At the same time, techniques have been proposed to intentionally increase the level of \gls{xt} to achieve the \gls{scr} when desired, as mentioned in \cref{sec:intro}. Concerning the \gls{imgcr}, it is in practice hard to ensure no inter-group coupling at all as it is hard to almost completely remove the impact of the perturbations. However, fiber designs with high $ \Delta $ tend to be more favorable in this sense \cite{sillard22}.}

Another relevant point to discuss is the assumption of weak-guidance. Since the weak-guidance approximation for which \gls{lp} modes exist is typically assumed in the literature for fibers with $ \Delta < 1\% $ \cite[p.38]{kawano01} \cite[p.20]{snyder83}, it might have been the case that such approximation did not hold for the \glspl{mmf} considered in this paper for which $ \Delta \in [1\%, 5\%] $. Thus, the accuracy of the numerical results has been checked for some points of Fig.\ref{fig:giCycle}, including some for which $ \Delta = 5\% $, generating the vector-modal profiles with the mode solver \hlbox{implemented in MATLAB as described in \cite{tasnadThesis} and based on \cite{fallahkhair08}}.\footnote{\hlbox{The mode solver described in \cite{fallahkhair08} is freely accessible as ``Thomas Murphy (2024). Waveguide Mode Solver (\url{https://www.mathworks.com/matlabcentral/fileexchange/12734-waveguide-mode-solver}), MATLAB Central File Exchange. Retrieved August 12, 2024.''}}
\hlbox{A square grid of $ N_\mathrm{p}\,\times\,N_\mathrm{p} $ evenly spaced points (with $ N_\mathrm{p} \ge 600$) has been used, covering a square integration area with edge length of $ \qty{125}{\mu m}$.}
For the fibers we tested, we observed a maximum discrepancy between the elements of $ \mat{\kappa} $ computed with the \gls{lp} and with the vector mode solver below $ 2\% $ (and much lower in terms of $ \kappa $ in the \gls{scr}), justifying the adoption of the \gls{lp} solver (mentioned in \cref{sec:fiberDesign}), which is computationally less expensive. The number of guided modes for each generated fiber in the weak-guidance has also been compared against the number of guided vector modes without such assumption \cite{hashimoto80}, finding negligible discrepancies.
\section{Conclusions}\label{sec:conclusions}
We have investigated the scaling of the nonlinear coupling coefficients $ \gamma \kappa $ and $ \gamma \mat{\kappa} $, appearing in the Manakov equations for the \gls{scr} and the \gls{imgcr}, respectively, with the number of modes $ M $.

\hlbox{Closed-form expressions have been derived, which depend on few fiber design parameters, namely, core radius $ R $, refractive index difference between core and cladding $ \Delta $, and number of guided modes. Validation has been performed with a numerical approach. The closed-form expressions provide a significantly quicker tool to compute $ \gamma \kappa $ and $ \gamma \mat{\kappa} $. Our analysis indicates that the elements of $ \mat{\kappa} $ are simple rational numbers which depend only on $ M $. As $ \gamma $ depends on the fundamental mode effective area $ \aeff{1}{1} $, which increases with $ R $ and reduces with $ \Delta $, maximizing $ \aeff{1}{1} $ minimizes the elements of $ \gamma \mat{\kappa} $, for a fixed $ M $.}

\hlbox{Even though the analysis has been carried out for trenchless parabolic \glspl{gimmf}, the comparison with values of $ \gamma \kappa $ computed for optimized and manufactured \glspl{mmf} confirms the validity of our approach for realistic fiber designs employing trenches and non-parabolic graded-index profiles.}

\hlbox{The insights developed in this paper can be helpful for fiber design}, and the closed-form expressions can be used, e.g.,  in nonlinear \gls{sdm} channel models \cite{serena22, lasagni23, garcia22}. Thus, contributing towards the assessment of the feasibility of a future long-haul \gls{sdm} communication system.
\appendices
\section{Closed-Form Expression for the Fundamental Effective Area of GIMMFs: Varying the Core Radius}\label{sec:aeff}
In order to obtain an approximate analytic expression for $ \aeff{1}{1} $ when varying only $ R $, let the fundamental mode be approximated with a Gaussian function as \cite[Eq. 15-2]{snyder83} \cite[Eq. 2.2.38]{agrawal21} \cite[Eq. 2.74]{keiser10}
\begin{equation}\label{key}
	c \, \exp{\Big(\frac{\rho}{r_0}\Big)^2}
\end{equation}
where $ c \ge 0 $ is some scaling factor, $ \rho $ is the radial coordinate, $ r_0 $ is the so-called spot-size. Then, the fundamental mode effective area becomes \cite[p. 32]{agrawal21} 
\begin{equation}\label{eq:aeff}
	\aeff{1}{1} = \pi r_0^2
\end{equation}
The spot-size $ r_0 $ can be related to $ V $ through different approaches. An empirically-fitted formula is \cite[Eq. 11]{marcuse78}
\begin{equation}\label{eq:vGif}
	\frac{r_0}{R} = \frac{A}{\sqrt{V}} + \frac{B}{V^{1.5}} + \frac{C}{V^6}
\end{equation}
with $ A = \sqrt{2} $, $ B \approx 0.372 $, and $ C \approx 26.773 $. For the case of a parabolic \gls{gimmf}, the last two terms are relevant only for low values of $ V $ (and thus $ M $). Hence, we neglect them for the sake of simplicity. A similar relation can be obtained with a variational approach, through which it has been shown that, for the infinite parabolic refractive index profile, \cite[Table 14-2]{snyder83}
\begin{equation}\label{key}
	\frac{r_0}{R} \propto \frac{1}{\sqrt{V}}
\end{equation}
where $ \propto $ means ``proportional to''.
A similar dependence has been obtained for the Gaussian refractive index profile \cite[Table 15-2]{snyder83}.

From \cref{eq:aeff} and \cref{eq:vGif}, it turns out that
\begin{equation}\label{eq:aeffVsV}
	\aeff{1}{1} \approx \pi \frac{2R^2}{V}
\end{equation}

Expressing $ R $ in terms of $ V $, substituting \cref{eq:nmodesgimmf} in \cref{eq:aeffVsV}, and remembering that $ \Delta $ is fixed  (i.e., $ M $ changes only because of $ R $), yields
\begin{equation*}
	\aeff{1}{1} \approx \pi \left(\frac{1}{\mathrm{NA} k_0}\right)^2 4 \sqrt{M}
\end{equation*}

\section{Closed-Form Expression for the Fundamental Effective Area of GIMMFs: Varying the Refractive Index Difference}\label{sec:aeffDelta}
The approximate closed formula for $ \aeff{1}{1} $ when $ R $ is fixed can be retrieved with the same Gaussian approach employed in Appendix \ref{sec:aeff}. With the help of  \cref{eq:aeffVsV} and \cref{eq:nmodesgimmf}, remembering that this time $ R $ is fixed and $ \Delta $ varies, it is found that 
\begin{equation*}
	\aeff11 \approx \frac{\pi R^2 A^2}{2} \frac{1}{\sqrt{M}} 
\end{equation*}

\section{Analytic Expression for the Manakov Nonlinearity Coefficients of GIMMFs}\label{sec:k}
In this section we derive approximate analytic results for the generic nonlinearity coefficient $ \kappa_{\alpha \beta} $ of the matrix  $ \mat{\kappa} $ for \glspl{gimmf} defined by \cref{eq:kSc}. We remind that $ \mat{\kappa} $ reduces to the scalar $ \kappa $ in case of a single group of strongly-coupled modes.

In order to derive an approximate analytic expression for $ \kappa_{\alpha \beta} $ for \glspl{gimmf}, the infinite parabolic profile approximation can be exploited, which provides relatively simple analytic expressions for the modal profiles, while for the actual parabolic profile of \cref{eq:gi} the analytic expression would be more complicated, involving the Whittaker functions of the first kind \cite[Sec. 14-8]{snyder83}.
This approach is commonly employed to obtain closed-form expressions for relevant quantities of \glspl{gimmf} like propagation constants and their relative cutoff values \cite[Sec.11.2.2]{agrawal21}, modal profiles, and nonlinear coefficients \cite{mafi12}. The approximation consists in assuming that the refractive index profile is parabolic everywhere, not only in the core. That is, it assumes that \cref{eq:gi} holds also for $ \rho>R $. The approximation is reasonable for low-order modes and a large enough core radius $ R $ \cite[Sec.11.2.2]{agrawal21}, even though we successfully exploit it for all modes and core radii later.

The modal profiles are then \cite[Table 14-2]{snyder83}\cite{mafi12}
\begin{equation}\label{eq:mode}
	\mode{p}{m}(\rho, \phi) = \moder{p}{m}(\rho) \modepol{m}(\phi)
\end{equation}
where $ p $ and $ m $ are two integers, $ \rho $ is the radial coordinate, $ \phi $ is the azimuth coordinate, and \cite{mafi12}
\begin{equation}\label{eq:moder}
	\moder{p}{m}(\rho) = A_p^m \frac{\rho^m}{\rho_0^{m + 1}} \expp{-\frac{\rho^2}{2\rho_0^2}} \lag{p}{m}\left(\frac{\rho^2}{\rho_0^2}\right)
\end{equation}
where $ \lag{p}{m} $ is a generalized Laguerre polynomial. The quantity $ \moder{p}{m}(\rho) $ accounts for the radial dependence of the modal profile with
\begin{align}\label{key}
	A_p^m &= \sqrt{\frac{p!}{\pi \, (p + m)!}}\\
	\rho_0 &= \frac{R}{(4N_2 )^{1/4}}\\
	N_2 &= \frac{1}{2} n_0^2 k_0^2 R^2 \Delta
\end{align}
The coefficient $ A_p^m $ has been chosen to fulfill the following orthonormality condition among the modes \cite[Eq.5]{mafi12}
\begin{equation}\label{eq:ortho}
	\int_{0}^{2\pi} \int_{0}^{\infty} \rho \mode{p'}{m'}^*(\rho, \phi) \mode{p}{m}(\rho, \phi) \dif \rho \dif \phi = \delta_{p, p'} \delta_{m,m'}
\end{equation}

The $ \lag{p}{m} $ are the generalized Laguerre polynomials which can be expressed in different ways. A handy approach for us is \cite[Eq. 18.59]{arfken11}
\begin{equation}\label{eq:lag}
	\lag{p}{m}(\rho) = \sum_{i = 0}^{n} b_i(n,m) \rho^i
\end{equation}
where 
\begin{equation}\label{key}
	b_i(n,m) = \left(-1\right)^i \binom{p + m}{p -  i} \frac{1}{i!}
\end{equation}
With simple algebraic passages we derived the following expression which can speed up the numerical implementation
\begin{equation}\label{key}
	b_{i+1}(p,m) = - \frac{p-i}{(i+1)(m+i+1)} b_i(p,m)
\end{equation}

The expression of the polarization and azimuth dependence $ \modepol{m}(\phi) $ appearing in \cref{eq:mode} depends on the chosen modal basis. Note that, in case of the LP modes, to draw a link with the classic notation, the mode $ \mode{p}{m} $ corresponds to $ \mathrm{LP}_{m,p+1} $\cite[p. 308]{snyder83}. In case of vector modes, within weak-guidance, the expressions for the polarization vectors $ \modepol{m}(\phi) $ are \cite[Eq. 1]{jiang17}
\begin{gather}
	p \ge 0 \notag \\
	\begin{cases}
		\mathrm{TM}_{0,p+1}(\phi) = \cos{(\phi)} \ver{x} + \sin{(\phi)} \ver{y} \\
		\mathrm{TE}_{0,p+1}(\phi) = \sin{(\phi)} \ver{x} - \cos{(\phi)} \ver{y}
	\end{cases}\label{eq:vecModes1}\\[1em]
	m > 1, p \ge 0 \notag\\
	\begin{cases}
		\mathrm{EH}_{m-1,p+1}^\mathrm{e}(\phi) = \cos{(m \phi)} \ver{x} + \sin{(m \phi)} \ver{y} \\
		\mathrm{EH}_{m-1,p+1}^\mathrm{o}(\phi) = \sin{(m \phi)} \ver{x} - \cos{(m \phi)} \ver{y}
	\end{cases}\label{eq:vecModes2}\\[1em]
	m \ge 1, p \ge 0 \notag\\
	\begin{cases}
		\mathrm{HE}_{m+1,p+1}^\mathrm{e}(\phi) = \cos{(m \phi)} \ver{x} - \sin{(m \phi)} \ver{y} \\
		\mathrm{HE}_{m+1,p+1}^\mathrm{o}(\phi) = \sin{(m \phi)} \ver{x} + \cos{(m \phi)} \ver{y}
	\end{cases}\label{eq:vecModes3}
\end{gather}
Note that all modal bases obtained as a unitary transformation of (quasi-)degenerate vector modes lead to the same $ \gamma \kappa_{\alpha \beta} $ \cite{schmidt17, antonelli19}. Hence, in principle the LP and vector modal basis are equivalent. However, we adopt the vector modes since they simplify the calculations later on.

Starting from Eqs. \ref{eq:kImgc}, \ref{eq:aeffInter}, and \ref{eq:aeffInterBar}, the generic nonlinearity coefficient $ \kappa_{\alpha \beta} $ can be computed as
\begin{equation}\label{eq:kImgcAlt}
	\kappa_{\alpha \beta} = \frac{4}{3} \frac{1}{(2N_\alpha + \delta_{\alpha,\beta})2N_\beta} \sum_{a\in I_\alpha}\sum_{b\in I_\beta} \aeff{1}{1}D_{ab}
\end{equation}
where we exploited the orthonormality so that $ \int_{-\infty}^{\infty}\int_{-\infty}^{\infty} \norm{\modeS{a}}^2 \dif x \dif y = \int_{-\infty}^{\infty}\int_{-\infty}^{\infty} \norm{\modeS{b}}^2 \dif x \dif y = 1$, and we defined
\begin{equation}\label{eq:DabDef}
	D_{ab} =\int_{0}^{\infty}\int_{0}^{2\pi} \rho \norm{\modeS{a}}^2 \norm{\modeS{b}}^2  \dif \phi \dif \rho 
\end{equation}
where $ a $ stands for the pair of mode indices $ (p, m) $ and $ b $ stands for $ (s,v) $. Hence, we now derive a closed-form expression for $ D_{ab} $.

Note that \cref{eq:kImgc} has been originally derived for LP modes only, while for vector modes a (slightly) more complex expression is normally used \cite[Eq. 56]{antonelli16}. We show in Appendix \ref{sec:simplerFormula} that \cref{eq:kImgc} holds also for vector modes (within weak-guidance) and hence we are allowed to consider \cref{eq:DabDef} for vector modes as well.\footnote{Since employing vector modes, it is enough to account for one mode per LP group, where an LP group consists of either $ 2 $ modes (for the $ \mathrm{LP}_{0x} $ modes (with $ x \in \{1, 2, \dots\} $)), or $ 4 $ modes. Indeed, $ D_{ab} $ depends only on the norm of the modes, which is the same for all vector modes within the same LP mode group (while this is not the case for the LP modes of a certain LP mode group). Hence, the later symbolic calculations are speeded up. Note that when only one vector mode per LP mode is employed, a suitable scaling factor within the summation of \eqref{eq:kImgcAlt} is needed.} Then,
\begin{align}\label{eq:kDen}
		D_{ab} &=\!\! \int_{0}^{\infty}\!\!\!\! \int_{0}^{2\pi}\!\!\!\!\!\! \rho \norm{\moder{p}{m}(\rho) \modepol{m}(\phi)}^2 \norm{\moder{s}{v}(\rho) \modepol{v}(\phi)}^2 \! \dif \phi \! \dif \rho \notag \\
		&= 2\pi \int_{0}^{\infty}\rho \norm{\moder{p}{m}(\rho) }^2 \norm{\moder{s}{v}(\rho)}^2 \dif \rho 
\end{align}
where we made use of $ \norm{\modepol{m}(\phi)}^2 = \norm{\modepol{v}(\phi)}^2 = 1 $, thanks to our choice of modal basis.

Substituting \cref{eq:moder} into \cref{eq:kDen}
\begin{multline}\label{key}
	D_{ab} = \tilde{C}_{ab} \int_{0}^{\infty} \rho \rho^{2m + 2v} \expp{-2\frac{\rho^2}{\rho_0^2}} \lag{p}{m}\left( \frac{\rho^2}{\rho_0^2} \right) \\
	\lag{p}{m}\left( \frac{\rho^2}{\rho_0^2} \right) \lag{s}{v}\left( \frac{\rho^2}{\rho_0^2} \right) \lag{s}{v}\left( \frac{\rho^2}{\rho_0^2} \right) \dif \rho
\end{multline}
where $ \tilde{C}_{pmsv} =  2\pi\left(\frac{A_p^m A_s^v}{\rho_0^{m+1}\rho_0^{v+1}}\right)^2 $ and $ \tilde{C}_{ab} $ is a short-hand notation for $ \tilde{C}_{pmsv} $.
We perform the change of variable $ u = \rho^2 $, so that $ \dif u = 2\rho \dif \rho $, which yields
\begin{multline}\label{eq:kDenInt}
	D_{ab} = C_{ab} \int_{0}^{\infty} u^{\tau_{mv}} \expp{-\sigma u} \lag{p}{m} \left( \frac{u}{\lambda} \right) \lag{p}{m}\left( \frac{u}{\lambda} \right) \\
	 \lag{s}{v} \left( \frac{u}{\lambda} \right) \lag{s}{v}\left( \frac{u}{\lambda} \right) \dif u
\end{multline}
where $ C_{ab} = \tilde{C}_{ab}/2 = \pi\left(\frac{A_p^m A_s^v}{\rho_0^{m+v+2}}\right)^2 $, $ \tau_{mv} = m+v $, $ \sigma = \rho_0^2/2 $, and $ \lambda = \rho_0^2 $. We now follow similar steps to \cite{miller63} to retrieve a closed-form expression for the previous formula. Plugging \cref{eq:lag} into \cref{eq:kDenInt}, it is obtained
\begin{multline}\label{eq:DInterm}
	D_{ab} = C_{ab} \sum_{i=0}^{p} \sum_{j=0}^{p} \sum_{k=0}^{s} \sum_{l=0}^{s} b_i(p,m) b_j(p,m)\\
	b_k(s,v) b_l(s,v) \lambda^{\epsilon-\tau_{mv}} \int_{0}^{\infty} \expp{-\sigma u} u^{\epsilon} \dif u
\end{multline}
where $\epsilon = i+j+k+l+\tau_{mv}$.
With the change of variable $ x = \sigma u $, the integral $ \int_{0}^{\infty} \expp{-\sigma u} u^{\epsilon}\dif u $ becomes
\begin{equation}\label{key}
	\sigma^{-(\epsilon + 1)} \int_{0}^{\infty} \expp{-x} x^\epsilon  \dif x = \sigma^{-\epsilon + 1} \Gamma(\epsilon+1) = \sigma^{-(\epsilon+1)}\epsilon!
\end{equation}
where the last passage holds only for integer values of $ \epsilon $, and where $ \Gamma(\epsilon+1) = \int_{0}^{\infty} \expp{-x} x^\epsilon  \dif x$ is the well-known gamma function. Observe also that $ C_{ab} \lambda^{\epsilon-\tau_{mv}} \sigma^{-(\epsilon+1)} = C_{ab} \frac{\rho_0^{2(1+\tau_{mv})}}{2^{\epsilon+1}} = \frac{\left(A_p^m A_s^v \right)^2}{\rho_0^2 2^{\epsilon+1}} $. Then, \cref{eq:DInterm} becomes
\begin{equation}\label{eq:DInterm2}
	D_{ab} = \frac{\left(A_p^m A_s^v \right)^2}{\rho_0^2} \tilde{D}_{ab}
\end{equation}
where
\begin{equation}
\tilde{D}_{ab} = \sum_{i=0}^{p} \sum_{j=0}^{p} \sum_{k=0}^{s} \sum_{l=0}^{s} b_i(p,m) b_j(p,m)
b_k(s,v) b_l(s,v) \frac{\epsilon!}{2^{\epsilon+1}}
\end{equation}
Inserting \cref{eq:DInterm2} in \cref{eq:kImgcAlt}, we obtain
\begin{equation}\label{eq:Dab}
	 \kappa_{\alpha \beta}
	 = \frac{4\pi}{3} \frac{1}{(2N_a+\delta_{ab})N_b} \sum_{a\in I_\alpha}\sum_{b\in I_\beta} \left(A_p^m A_s^v \right)^2 \tilde{D}_{ab}
\end{equation}
where we used $ \aeff{1}{1} = 2\pi\rho_0^2 $ \cite{mafi12}.

Notice from \cref{eq:Dab} that $ \kappa_{\alpha \beta} $ does not depend on $ R $ or $ \Delta $, but just on the number of modes, and it is expressed only in terms of ratios of integers. Hence, $ \kappa_{\alpha \beta} $ can be computed exactly (within the infinite parabolic approximation) with, e.g., the help of a software with symbolic computation capabilities, like MATLAB, as we did. To know which set of modal indices $ I_\alpha $ corresponds to a certain mode group, the approach in \cite{lukowski77} is helpful. \hlbox{The code we wrote to carry out the symbolic calculations for \cref{eq:Dab} has been uploaded to \cite{mmfData}.} The analytic values of $ \kappa $ in the \gls{scr} for up to $ 32 $ mode groups are reported in \cref{tab:kSc}. For the \gls{imgcr} the analytic values are reported in \cref{tab:kImgc} for up to $ 10 $ mode groups ($ 110 $ modes), and in \cite{mmfData} for up to $ 32 $ mode groups ($ 1056 $ modes).

\section{Intermodal Effective Area for Vector Modes}\label{sec:simplerFormula}
In this section we prove the claim made in Appendix \ref{sec:k} that \cref{eq:kImgc} holds not only for the LP modal basis as per \cite{antonelli16}, but for vector modes \ref{eq:vecModes1}-\ref{eq:vecModes3} as well. For a basis of real modes in the weakly guiding regime, like the vector modal basis, the nonlinear coupling coefficients $ \gamma \kappa_{ab} $ can be computed as \cite[Eq. 55]{antonelli16}
\begin{equation}\label{eq:kDefGen}
	\gamma \kappa_{\alpha\beta} = \frac{\omega_0 n_2}{c Z_0^2} \frac{1}{(M_\alpha + \delta_{\alpha\beta}) M_\beta } \sum_{a\in I_\alpha} \sum_{b \in I_\beta} \frac{n_\mathrm{eff}^2}{N_a^2 N_b^2} I_{ab}
\end{equation}
where 
\begin{equation}\label{key}
	I_{ab} = \! \frac{1}{6} \int_{-\infty}^{\infty} \!\! \int_{-\infty}^{\infty} \!\!\! \left(\norm{\vec{F}_a}^2 \norm{\vec{F}_b}^2 + 2\norm{\vec{F}_a\cdot \vec{F}_b}^2\right) \dif x \dif y 
\end{equation}
\hlbox{Consider the two modes $ \modeS{a}$ and $ \modeS{a'} $ with same spatial profile $ \norm{\modeS{a}} = \norm{\modeS{a'}} $ and orthogonal polarizations, i.e., $ \modepol{a}(\phi) \cdot \modepol{a'}(\phi) = 0 $, and similarly for the modes $ \modeS{b}$ and  $\modeS{b'} $. Such two pair of modes can always be found for the vector modal basis by its definition \ref{eq:vecModes1}-\ref{eq:vecModes3}. A necessary and sufficient condition to obtain \cref{eq:kImgc} from \cref{eq:kDefGen} (besides the scaling factor $ \gamma $) is \cite[Eq. 57]{antonelli16}}
\begin{equation}\label{eq:toProve}
	I_{ab} + I_{a'b} + I_{ab'} + I_{a'b'} = \frac{4}{3} \int_{-\infty}^{\infty} \int_{-\infty}^{\infty} \norm{\modeS{a}}^2 \norm{\vec{F}_b}^2  \dif x \dif y 
\end{equation}

To prove that \cref{eq:toProve} holds, let us define
\begin{equation}\label{eq:I1}
	I_{1,ab} = \int_{-\infty}^{\infty} \int_{-\infty}^{\infty} \norm{\vec{F}_a}^2 \norm{\vec{F}_b}^2  \dif x \dif y 
\end{equation}
and 
\begin{equation}\label{eq:I2}
	I_{2,ab} = \int_{-\infty}^{\infty} \int_{-\infty}^{\infty} \norm{\vec{F}_a \cdot \vec{F}_b}^2 \dif x \dif y 
\end{equation}

\hlbox{Firstly, given that by assumption $ \norm{\modeS{a}} = \norm{\modeS{a'}} $ and $ \norm{\modeS{b}} = \norm{\modeS{b'}} $, we have that}
\begin{equation}\label{eq:s1}
	\hlbox{I_{1,ab} = I_{1,a'b} = I_{1,ab'} = I_{1,a'b'}}
\end{equation}
\hlbox{Given the separability of the radial and the azimuth dependence, i.e., the radial dependence can be factored out from the previous integrals, in the following we focus only on the azimuth dependence. We also make use of the fact that a polarization vector has unit norm by definition, i.e.,  $ \norm{\modepol{i}} = 1 $.}

\hlbox{Concerning $ I_{2,ab} $, remember that $ \modepol{a}\cdot\modepol{b} = \norm{\modepol{a}} \norm{\modepol{b}}\cos{(\theta)}$ $= \cos{(\theta)}$, with $ \theta $ being the angle between $ \modepol{a} $ and $ \modepol{b} $. Since $ \modepol{a'} $ is orthogonal to $ \modepol{a} $, the angle between $ \modepol{a'} $ and $ \modepol{b} $ is $ \pi/2 - \theta $. Hence, $ \modepol{a'} \cdot \modepol{b} = \sin{(\theta)} $. Given that $ \int_{0}^{2\pi} \sin^2{(\theta)} \dif\phi = \int_{0}^{2\pi} \cos^2{(\theta)} \dif\phi = \pi$, it follows that $ I_{2,ab} = I_{2,a'b} $. Additionally, due to the orthogonality between $ \modepol{a} $ and $ \modepol{a'} $, and between $ \modepol{b} $ and $ \modepol{b'} $, it follows that $ I_{2,ab} = I_{2,a'b'} $ and $ I_{2,a'b} = I_{2,ab'} $. Hence,}
\begin{equation}\label{eq:s2}
	\hlbox{I_{2,ab} = I_{2,a'b} = I_{2,ab'} = I_{2,a'b'}}
\end{equation} 
\hlbox{Note also that}
\[  \hlbox{I_{2,ab} \propto \int_{0}^{2\pi} \norm{\modepol{a}(\phi)\cdot\modepol{b}(\phi)}^2 \dif \phi = \int_{0}^{2\pi} \cos^2{(\theta)} \dif \phi = \pi}  \]
and 
\[ \hlbox{I_{1,ab} \propto \int_{0}^{2\pi} \norm{\modepol{a}(\phi)}^2 \norm{\modepol{b}(\phi)}^2 \dif \phi = 2\pi} \]
Thus, 
\begin{equation}\label{eq:s3}
	\hlbox{I_{2,ab} = I_{1,ab}/2}
\end{equation}

\hlbox{Hence, plugging \cref{eq:s1} and \cref{eq:s2} into the left-hand side of \cref{eq:toProve}, and using \cref{eq:s3}, we get $ I_{ab} + I_{a'b} + I_{ab'} + I_{a'b'} = 4/6 I_{1,ab} + 8/6 I_{2,ab} = 4/6 I_{1,ab} + 4/6 I_{1,ab} = 4/3 I_{1,ab} $, which proves \cref{eq:toProve}.}

\section{Nonlinearity Coefficients of Step-Index Fibers}\label{sec:simmfExtra}
The derivation of closed-form expressions for the nonlinear coupling coefficients in \glspl{simmf} has few differences compared to \glspl{gimmf}. Firstly, the dependence of $ \gamma $ on $ R $ and on $ \Delta $ is not the same as for \glspl{gimmf}. This is because for a \gls{simmf}, instead of \cref{eq:vGif}, the fitted relation for the spot-size of the Gaussian approximation is \cite[Eq. 8]{marcuse77}
\begin{equation}\label{eq:gApprox1}
	\frac{r_0}{R} = A + \frac{B}{V^{1.5}} + \frac{C}{V^6}
\end{equation}
where $ A = 0.65 $, $ B = 1.619 $, $ C = 2.879 $, and we have verified the last term to be negligible in our case. 
The variational approach formula for a \gls{simmf} yields \cite[Table 15-2]{snyder83}
\begin{equation}\label{key}
	\frac{r_0}{R} \propto \frac{1}{\sqrt{\log{V}}}
\end{equation}
which, however, we have verified to be less accurate than \cref{eq:gApprox1}. Hence, when $ R $ is varied fixed $ \Delta $, with the help of Eqs.~\eqref{eq:aeff}, \eqref{eq:gApprox1} and \eqref{eq:nmodessimmf} it is found that
\begin{equation}\label{eq:siAeffScaling}
	\aeff{1}{1} \approx \pi \left(\frac{A}{\mathrm{NA} k_0}\right)^2 2 M \left(A + \frac{B}{(2M)^{0.75}} \right)^2
\end{equation}
If $ \Delta $ is varied, while keeping $ R $ fixed, then
\begin{equation}\label{eq:siAeffScaling2}
	\aeff{1}{1} \approx \pi R^2 \left(A + \frac{B}{(2M)^{0.75}} \right)^2
\end{equation}
As a side note, $ \aeff{1}{1} $ increases slower than in a \gls{gimmf} when $ R $ is fixed, and decreases slower when $ \Delta $ is fixed, due to the different relations between $ r_0 $ and $ V $.

We observed the numerical results to agree well with the theoretical scaling laws $ M \, \left(A + \frac{B}{(2M)^{0.75}} \right)^2 $ and $ \left(A + \frac{B}{(2M)^{0.75}} \right)^2 $, see Fig.\ref{fig:siAeff}, but for slightly different proportionality factors than $ \pi \left(\frac{A}{\mathrm{NA} k_0}\right)^2 \,  2 $ and $ \pi R^2 $.

The dependence of $ \kappa $ on $ M $ is numerically found to be essentially constant (thus weaker than $ 1/\sqrt{M} $ for \glspl{gimmf}), see Fig.\ref{fig:siKnl}, even though we did not provide an analytic derivation of $ \kappa $ as for \glspl{gimmf}. Ideally, it would be possible to consider the weakly guiding analytic expressions for the modal profiles of \glspl{simmf}, which depend on Bessel functions \cite[Table 14-6]{snyder83}, and follow a procedure analogous to the one for \glspl{gimmf} in Appendix \ref{sec:k}. For this purpose, formulas for integrals involving squares of Bessels functions (and their products) given in \cite[p. 431]{rosenheinrich22} and in \cite[Eq. 14]{matagne22} could be useful.

\section{Nonlinearity Coefficient for Non-Parabolic Graded Index Fibers}\label{sec:realFibersApp}
The deviations in $ \gamma\kappa $ in case of non-parabolic \glspl{gimmf} could be ideally studied  with a similar approach to the one in \cref{sec:scaling}. That is, by independently approximating $ \gamma $  (or, equivalently, $ \aeff{1}{1} $) and $ \kappa $. Strictly speaking, it would be necessary to recompute also the cutoff frequencies and, hence, a substitute relation for \cref{eq:nmodesgimmf}. However, given the grading exponent $ g = 2 + \delta $, with $ \delta \in [-10\,\%, +10\%] $ (conservative realistic range \cite{ferreira14}), it can be safely assumed that \cref{eq:nmodesgimmf} is almost untouched.

Concerning $ \kappa $, an analytic derivation in the sense of Appendix \ref{sec:k} would be quite involved because of the non-trivial expressions of the modal profiles. Extensive computations with the numerical method could be carried out to fit a relation similar to \cref{eq:kImgcPaolo} which relates $ \kappa_{ab} $ to $ M $. However, we do not expect significant changes in view of a little $ \delta $, as confirmed by \cref{fig:giCycleWithFilipe}, for which we verified that the values of $ \kappa $ numerically computed for fibers in the literature (which in general do not have a parabolic graded index) are close to the ones of parabolic \glspl{gimmf}.

For $ \aeff{1}{1} $, it is possible to exploit the Gaussian approximation for a generic \gls{gi} profile \cite[Eq. 11]{marcuse78}. 
Then, with the same procedure as in Appendices \ref{sec:aeff} and \ref{sec:aeffDelta}, for a generic power-law \gls{gi} profile with grading exponent $ g = 2+ \delta $, neglecting the last two terms of the Gaussian approximation for $ r_0 / R $ \cite[Eq. 11]{marcuse78}, it holds
\begin{equation}\label{eq:giMyFormulaGen}
	\gamma\,\kappa (\delta) \, \approx \frac{n_2 \omega_0}{c} \frac{7}{2\pi R^2} \frac{V^{\frac{-\delta}{4+\delta}}}{\bar{A}^2} 
\end{equation}
where $ \bar{A} = (\frac{2}{5}(1+4(\frac{2}{2+\delta})^{5/6}))^{0.5} $.

Then, the ratio between $ \gamma\,\kappa $ of \cref{eq:giMyFormulaGen} and the parabolic case of \cref{eq:giMyFormula2} is given by
\begin{equation}\label{key}
	\epsilon = \frac{ \gamma\kappa \, (\delta)}{\gamma\kappa \, (\delta = 0) } = V^{\frac{-\delta}{4+\delta}}\frac{2}{\bar{A}^2}
\end{equation}
For $ V \in [1, 100] $, which covers the whole range of Fig.\ref{fig:giCycleWithFilipe}, and $ \delta \in [-10\,\%, +10\%] $, it has been numerically observed that the maximum excursion of $ \epsilon $ is $ \pm 10\% $.

\section*{Acknowledgments}
\noindent The authors acknowledge fruitful discussions with Paolo Serena of the University of Parma.

\bibliographystyle{IEEEtran}
\bibliography{parts/bib.bib}

\vspace{11pt}

\begin{IEEEbiography}[{\includegraphics[width=1in,height=1.25in,clip,keepaspectratio]{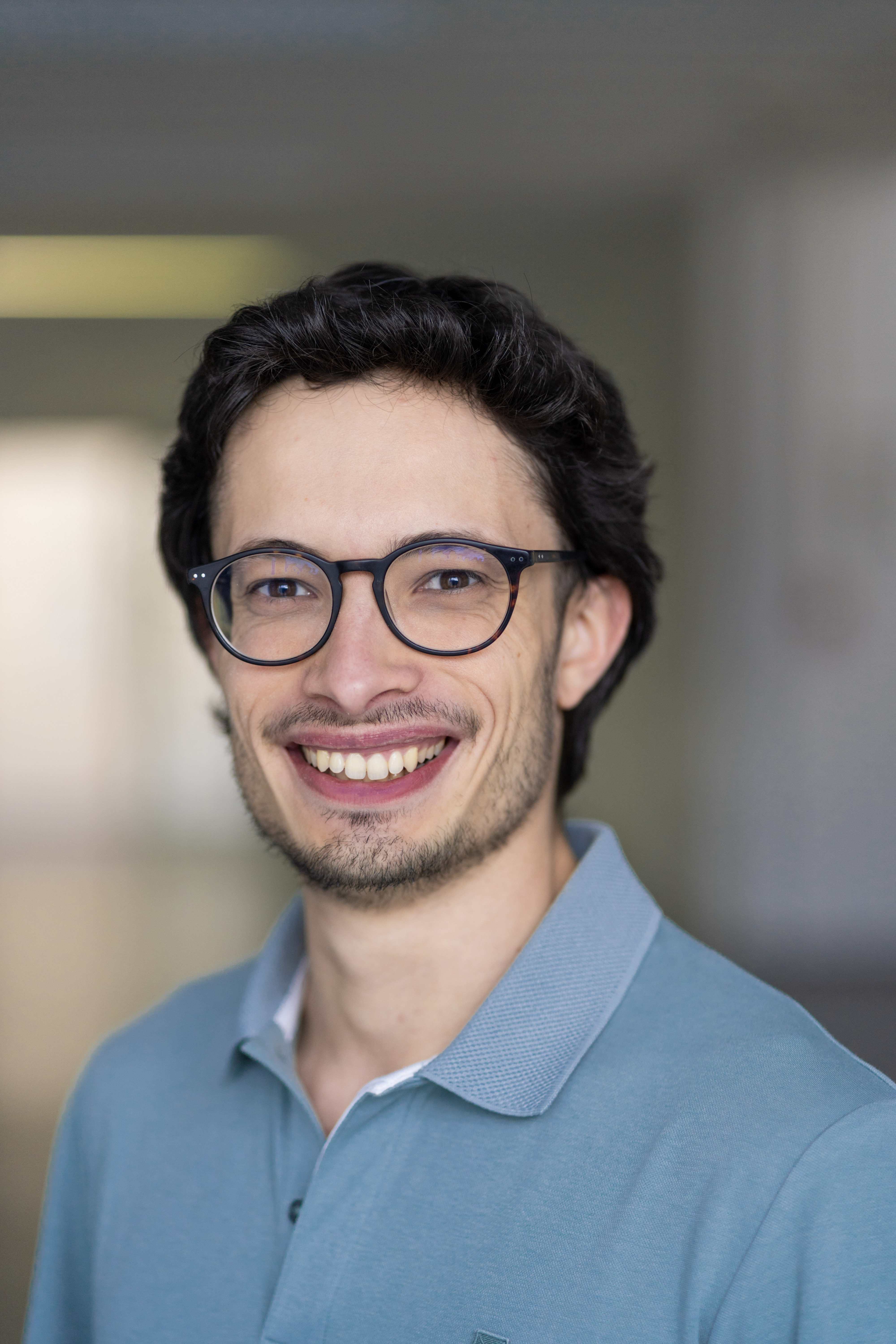}}]{Paolo Carniello}
	received the B.S. degree from the University of Udine, Italy, in electronic engineering in 2019, and the M.S. degree from Politecnico di Torino, Italy, in telecommunications engineering in 2021. He is currently working towards the Ph.D. degree at the Technical University of Munich, Germany. His research interests focus on space-division multiplexing for optical communications systems: channel modeling, signal processing, and aspects of applied information theory.
\end{IEEEbiography}

\begin{IEEEbiography}[{\includegraphics[width=1in,height=1.25in,clip,keepaspectratio]{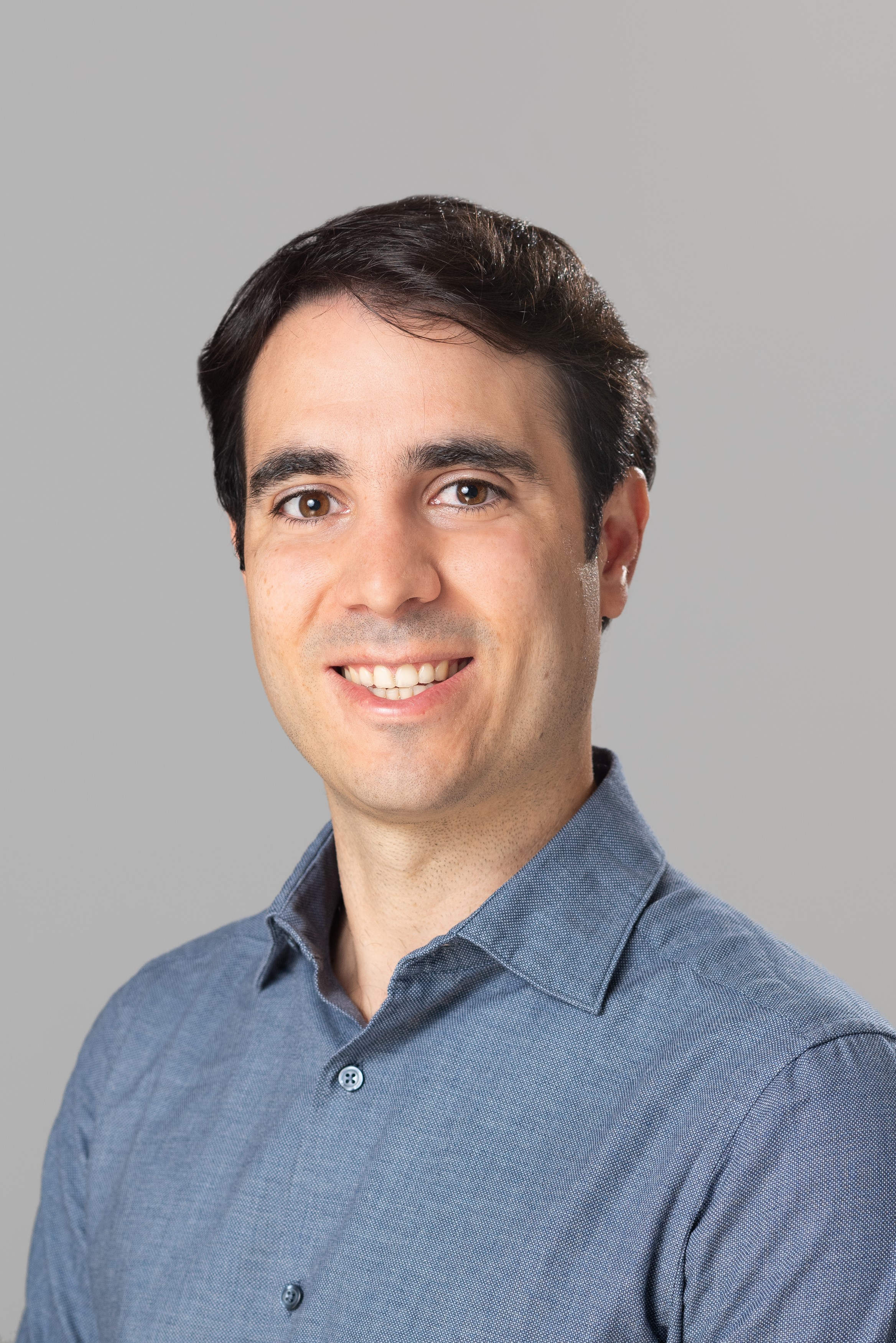}}]{Filipe Ferreira}
	 is a Principal Research Fellow and UKRI Future Leaders Fellow in the Department of Electronic and Electrical Engineering at University College London. His research focuses on exploring massive spatial parallelism in optical fibres by integrating spatial light modulation with machine learning. Filipe also serves as the Chair of the IEEE UK \& Ireland Photonics Society Chapter, contributing to the advancement of the photonics community. He earned his PhD in Electrical and Computer Engineering from the University of Coimbra in 2014, specializing in high-capacity optical transmission in multipath fibres to surpass the limits of conventional single-mode systems. During his doctoral studies, Filipe worked as a research engineer at Nokia Siemens Networks, where he led advancements in mode-division multiplexing as part of the EU-FP7 project MODEGAP. Subsequently, he was awarded a prestigious Marie Skłodowska-Curie Fellowship to continue his research at Aston University, where he made significant contributions to understanding the interplay between linear and nonlinear impairments in multi-mode fibres and their impact on transmission capacity. In 2020, Filipe began his UKRI Future Leaders Fellowship at UCL, focusing on high-capacity optical fibre transmission, spatially multiplexed systems, digital nonlinear compensation, and advanced optical fibre design. He has co-authored over 140 peer-reviewed journal and conference papers, including numerous invited contributions, and delivered more than 30 invited talks at major international events such as ECOC and OFC. He is also the lead inventor on four active WO patents.
\end{IEEEbiography}

\begin{IEEEbiography}[{\includegraphics[width=1in,height=1.25in,clip,keepaspectratio]{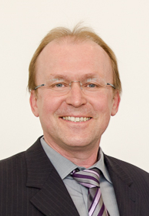}}]{Norbert Hanik}
	received the Dipl.-Ing. degree in electrical engineering (with a thesis on Digital Spread Spectrum Systems) and the Dr.-Ing. degree (with a dissertation on nonlinear effects in optical signal transmission) from the Munich University of Technology, Munich, Germany, in 1989 and 1995, respectively. He was a Research Associate at the Institute for Telecommunications, Technical University of Munich, from 1989 to 1995, where he worked in the area of Optical Communications. From 1995 to 2004 he has been with the Technology Center of Deutsche Telekom, Berlin, heading the research group System Concepts of Photonic Networks. During his work there, he contributed to a multitude of national, international, and Telekom internal R\&D-projects. From January to March 2002, he was a Visiting Professor at Research Center COM, Technical University of Denmark, Lingby. Since 2004 he holds the Professorship for Line Transmission Technology at the Technical University of Munich. His primary research interests are in the fields of design and optimization of optical communication systems.
\end{IEEEbiography}

\end{document}